\documentclass{llncs}
\usepackage{amssymb}
\usepackage{amsmath}
\usepackage{graphicx}

\input tcilatex
\begin{document}

\author{Zining Cao \\
%EndAName
\institute{
          College of Computer Science and Technology \\ Nanjing University of Aeronautics and Astronautics, Nanjing 211106, China \\
    \email{caozn@nuaa.edu.cn}}}
\title{A Complete Mental Temporal Logic for Intelligent Agents}
\maketitle

\begin{abstract}
In this paper, we present a mental temporal logic \textit{BPICTL} together
with a sound and complete axiomatization, which generalizes \textit{CTL} by
introducing mental modalities$.$ A sound and complete inference system of 
\textit{BPICTL} is given. We also prove the finite model property and
decidability of \textit{BPICTL}. Furthermore, we present a model checking
algorithm for \textit{BPICTL}. \newline
\textbf{Key Words.} {Mental Temporal Logic; Soundness and Completeness;
Model Checking}
\end{abstract}

\section{Introduction}

The field of multi-agent systems (\textit{MAS}) theories is traditionally
concerned with the formal representation of the mental attitudes of
autonomous entities, or agents, in a distributed system. For this task
several modal logics have been developed. The most studied logics include
logics for knowledge, beliefs, desires, goals, and intentions. These logics
are seen as specifications of particular classes of \textit{MAS} systems.
Their aim is to offer a description of the collective mental attitudes (such
as knowledge, belief and etc.) that a \textit{MAS} should exhibit in a
specific class of scenarios. A considerable number of these formal studies
are available in the literature and temporal extensions of these (i.e.,
modal combinations of \textit{CTL} or \textit{LTL} with the modalities for
the mental attitudes) have been proposed \cite{DHK05,HW02,KLP04}. The
typical technical contribution of this line of work is to explore the
metalogical properties of these logics, e.g., completeness, decidability,
and computational complexity.

Verification of reaction systems by means of model checking techniques is a
well-established area of research \cite{CGP99}. In this paradigm one
typically models a system $S$ in terms of automata (or by a similar
transition-based formalism), builds an implementation $P_{S}$ of the system
by means of a model-checker friendly language such as the input for \textit{%
SMV} or \textit{PROMELA}, and finally uses a model-checker such as \textit{%
SMV} or \textit{SPIN} to verify some property $\varphi $ in the model $M_{P}$%
: $M_{P}\models \varphi ,$ where $M_{P}$ is a model representing the
executions of $P_{S}.$ The field of multi-agent systems has also become
interested in the problem of verifying complex systems \cite{BB05,WFHP02}.
In \textit{MAS}, modal logics represent concepts such as knowledge, belief,
and intention. Since these modalities are given interpretations that are
different from the ones of the standard temporal operators, it is not
straightforward to apply existing model checking tools developed for \textit{%
LTL\TEXTsymbol{\backslash}CTL} temporal logic to the specification of 
\textit{MAS}$.$

In this paper, we present a mental temporal logic $BPICTL,$ which is an
extension of $CTL$ by adding belief modality, preference modality and
intention modality$.$ Some new interaction properties between mental
modalities and temporal modalities are given. We prove the soundness,
completeness and finite model property of $BPICTL$. Furthermore we present a
model checking algorithm for $BPICTL$. The rest of the paper is organized as
follows: In Section 2, we present a mental temporal logic $BPICTL,$ give its
syntax, semantics and inference system. In Section 3, we study the
soundness, completeness. In Section 4, finite model property and
decidability of $BPICTL$ are given$.$ In Section 5, the approach to model
checking $BPICTL$ is studied. The paper is concluded in Section 6.

\section{A Mental Temporal Logic $BPICTL$}

In the research of \textit{MAS, }several mental logics were studied to
describe the mental attitudes of agents. For example, \textit{BDI} logic 
\cite{RG91} is a famous mental logic that specifies beliefs, desires and
intentions of agents. On the other hand, several temporal logics were
presented to specify temporal property of systems such as \textit{CTL }\cite%
{CE81}. To reason mental and temporal properties in \textit{MAS},\textit{\ }%
we present a mental temporal logic $BPICTL$ in this section, which can be
viewed as an extension of \textit{BDI} logic and \textit{CTL} logic.

\subsection{Syntax of $BPICTL$}

Throughout this paper, we let $L^{BPICTL}$ be a language which is just the
set of formulas of interest to us.

\textbf{Definition 1.} The set of formulas in $BPICTL$, called $L^{BPICTL}$,
is given by the following rules:

(1) If $\varphi \in $atomic formulas set $\Pi $, then $\varphi \in
L^{BPICTL} $.

(2) If $\varphi \in L^{BPICTL}$, then $\lnot \varphi \in L^{BPICTL}$.

(3) If $\varphi _{1}$, $\varphi _{2}\in L^{BPICTL}$, then $\varphi
_{1}\wedge \varphi _{2}\in L^{BPICTL}$.

(4) If $\varphi _{1}$, $\varphi _{2}\in L^{BPICTL}$, then $\varphi _{1}\vee
\varphi _{2}\in L^{BPICTL}$.

(5) If $\varphi \in L^{BPICTL}$, then $B_{a}\varphi \in L^{BPICTL}$.
Intuitively, $B_{a}\varphi $ means that agent $a$ believes $\varphi $.

(6) If $\varphi \in L^{BPICTL}$, then $P_{a}\varphi \in L^{BPICTL}$.
Intuitively, $P_{a}\varphi $ means that agent $a$ prefers $\varphi $.

(7) If $\varphi \in L^{BPICTL}$, then $I_{a}\varphi \in L^{BPICTL}$.
Intuitively, $I_{a}\varphi $ means that agent $a$ intends $\varphi $.

(8) If $\varphi \in L^{BPICTL}$, then $AX\varphi \in L^{BPICTL}$.
Intuitively, $AX\varphi $ means that every next state satisfies $\varphi .$

(9) If $\varphi \in L^{BPICTL}$, then $EX\varphi \in L^{BPICTL}$.
Intuitively, $EX\varphi $ means that there is a next state which satisfies $%
\varphi .$

(10) If $\varphi \in L^{BPICTL}$, then $EF\varphi \in L^{BPICTL}$.
Intuitively, $EF\varphi $ means that property $\varphi $ holds at some state
on some path from the current state.

(11) If $\varphi \in L^{BPICTL}$, then $EG\varphi \in L^{BPICTL}$.
Intuitively, $EG\varphi $ means that property $\varphi $ holds at every
state on some path from the current state.

(12) If $\varphi _{1}$, $\varphi _{2}\in L^{BPICTL}$, then $E\varphi
_{1}U\varphi _{2}\in L^{BPICTL}$. Intuitively, $E\varphi _{1}U\varphi _{2}$
means that there exists a path from the current state, such that there is a
state on the path where $\varphi _{2}$ holds, and at every preceding state
on the path, $\varphi _{1}$ holds.

We use the following abbreviations:

$\varphi _{1}\rightarrow \varphi _{2}\overset{def}{=}\lnot \varphi _{1}\vee
\varphi _{2}.$

$\varphi _{1}\leftrightarrow \varphi _{2}\overset{def}{=}(\varphi
_{1}\rightarrow \varphi _{2})\wedge (\varphi _{2}\rightarrow \varphi _{1}).$

$Des_{a}\varphi \overset{def}{=}P_{a}\varphi \wedge B_{a}\lnot \varphi .$
Intuitively, $Des_{a}\varphi $ means that agent $a$ desires $\varphi $.

$AG\varphi \overset{def}{=}\lnot EF\lnot \varphi .$ Intuitively, $AG\varphi $
means that property $\varphi $ holds at every state on every path from the
current state.

$A\varphi _{1}U\varphi _{2}\overset{def}{=}\lnot E(\lnot \varphi _{2}U(\lnot
\varphi _{1}\wedge \lnot \varphi _{2}))\wedge \lnot EG\lnot \varphi _{2}.$
Intuitively, $A\varphi _{1}U\varphi _{2}$ means that for every path from the
current state, such that there is a state on the path where $\varphi _{2}$
holds, and at every preceding state on the path, $\varphi _{1}$ holds.

$AF\varphi \overset{def}{=}AtrueU\varphi .$ Intuitively, $AF\varphi $ means
that property $\varphi $ holds at some state on every path from the current
state.

Since $Des_{a}\varphi $ is represented in $BPICTL,$ \textit{BDI} logic can
be viewed as a sublogic of $BPICTL.$

\subsection{Semantics of $BPICTL$}

We will describe the semantics of $BPICTL$, that is, a formal model that we
can use to determine whether a given formula is true or false. Semantics of
normal modal logics are usually based on Kripke model. In $BPICTL,$
preference modality and intention modality are not normal modal operator, so
we give semantics of preference modality and intention modality based on
neighbourhood model \cite{Pac17}.

\textbf{Definition 2.} A model $M$ of $BPICTL$ is a structure $M=(S,$ $\Pi ,$
$\pi ,$ $R_{a}^{B},$ $R_{a}^{P},$ $R_{a}^{I},$ $R_{a}^{X},$ $D_{a})$, where

(1) $S$ is a nonempty set, whose elements are called possible worlds or
states.

(2) $\Pi $ is a nonempty set of atomic formulas.

(3) $\pi $ is a map: $S\times \Pi \longrightarrow \{true,false\}$.

(4) $R_{a}^{B}\subseteq S\times S$ is an accessibility relation on $S$,
which is a belief relation.

(5) $R_{a}^{P}:S\longrightarrow \wp (\wp (S))$, where $\wp (S)$ is the power
set of $S$.

(6) $R_{a}^{I}:S\longrightarrow \wp (\wp (S)).$

(7) $R_{a}^{X}\subseteq S\times S$ is an accessible relation on $S$, which
is a temporal relation.

(8) $D_{a}$ is a nonempty subset of $\wp (S)$, which satisfies the following
conditions:

(a) If $p$ is an atomic formula, then $Ev(p)=\{s$ $|$ $\pi (s,p)=true\}\in
D_{a}.$

(b) If $A\in D_{a}$, then $S-A\in D_{a}$.

(c) If $A_{1},A_{2}\in D_{a}$, then $A_{1}\cap A_{2}\in D_{a}$.

(d) If $A\in D_{a}$, then $Ev_{a}^{B}(A)\in D_{a}$. where $Ev_{a}^{B}(A)=\{s$
$|$ $s\in S$ such that $(s,t)\in R_{a}^{B}\Rightarrow t\in A\}.$

(e) If $A\in D_{a}$, then $Ev_{a}^{P}(A)\in D_{a}$. where $Ev_{a}^{P}(A)=\{s$
$|$ $s\in S$ such that $A\in R_{a}^{P}(s)\}.$

(f) If $A\in D_{a}$, then $Ev_{a}^{I}(A)\in D_{a}$. where $Ev_{a}^{I}(A)=\{s$
$|$ $s\in S$ such that $A\in R_{a}^{I}(s)\}.$

(g) If $A\in D_{a}$, then $Ev_{a}^{EX}(A)\in D_{a}$. where $%
Ev_{a}^{EX}(A)=\{s$ $|$ there exists $t\in A$ and $(s,t)\in R_{a}^{X}\}.$

(h) If $A\in D_{a}$, then $Ev_{a}^{EG}(A)\in D_{a}.$where $%
Ev_{a}^{EG}(A)=\{s $ $|$ there exists infinite trace $\rho =t_{0},t_{1},...$
from state $s=t_{0},$ where $(t_{i},t_{i+1})\in R_{a}^{X},$ such that $%
t_{i}\in A$ for all $i\}$.

(i) If $A_{1},A_{2}\in D_{a}$, then $Ev_{a}^{EU}(A_{1},A_{2})\in D_{a}.$%
where $Ev_{a}^{EU}(A_{1},A_{2})=\{s$ $|$ there exists infinite trace $\rho
=t_{0},t_{1},...$ from state $s=t_{0},$ where $(t_{i},t_{i+1})\in R_{a}^{X},$
and there exists a position $m\geq 0,$ such that $t_{m}\in A_{2}$ and for
all positions $0\leq k<m,$ we have $t_{k}\in A_{1}.\}.$

In order to give the semantics of $BPICTL,$ we define conditions of $%
R_{a}^{B},R_{a}^{P},R_{a}^{I},R_{a}^{X}$.

(B3) $(\forall x.\forall y.\forall z.((x,y)\in R_{a}^{B}\wedge (y,z)\in
R_{a}^{B}\rightarrow (x,z)\in R_{a}^{B}))$

(B4) $(\forall x.\forall y.\forall z.((x,y)\in R_{a}^{B}\wedge (x,z)\in
R_{a}^{B}\rightarrow (y,z)\in R_{a}^{B}))$

(B5) $(\forall x.\exists y.(x,y)\in R_{a}^{B})$

(P1) $\forall x.\forall Q_{1},Q_{2}\in D_{a}.(Q_{1}\in R_{a}^{P}(x)\wedge
Q_{2}\in R_{a}^{P}(x)\rightarrow Q_{1}\cap Q_{2}\in R_{a}^{P}(x))$

(P2) $\forall x.\forall Q_{1},Q_{2}\in D_{a}.(Q_{1}\in R_{a}^{P}(x)\wedge (%
\overline{Q_{1}}\cup Q_{2})\in R_{a}^{P}(x))\rightarrow Q_{2}\in
R_{a}^{P}(x),$ where $\overline{Q}$ is the complement of $Q$.

(P3) $\forall x.\forall Q.(\{y$ $|$ $Q\in R_{a}^{P}(y)\}\in
R_{a}^{P}(x))\rightarrow Q\in R_{a}^{P}(x)$

(P4) $\forall x.\forall Q_{1}.Q_{1}\in R_{a}^{P}(x)\rightarrow \forall
Q_{2}.\exists y.(Q_{2}\in R_{a}^{P}(x)\rightarrow (y\in Q_{2}\wedge Q_{1}\in
R_{a}^{P}(y)))$

(X1) $\forall x.\exists y.(x,y)\in R_{a}^{X}$

(BP1) $\forall x.\forall Q_{1},Q_{2}\in D_{a}.(Q_{1}\in R_{a}^{P}(x)\wedge
\{y$ $|$ $(x,y)\in R_{a}^{B}\}\subseteq (Q_{1}\cap Q_{2})\cup (\overline{%
Q_{1}}\cap \overline{Q_{2}}))\rightarrow Q_{2}\in R_{a}^{P}(x)$

(BP2) $\forall x.\forall Q.Q\in R_{a}^{P}(x)\rightarrow (\forall y.(x,y)\in
R_{a}^{B}\rightarrow Q\in R_{a}^{P}(y))$

(BP3) $\forall x.\forall Q\in D_{a}.(\exists y.(x,y)\in R_{a}^{B}\wedge Q\in
R_{a}^{P}(y))\rightarrow Q\in R_{a}^{P}(x)$

(BP4) $\forall x.\forall Q\in D_{a}.(\forall y.(x,y)\in R_{a}^{B}\rightarrow
Q\in R_{a}^{P}(y))\rightarrow Q\in R_{a}^{P}(x)$

(BP5) $\forall x.\forall Q\in D_{a}.Q\in R_{a}^{P}(x)\rightarrow \exists
y.((x,y)\in R_{a}^{B}\wedge Q\in R_{a}^{P}(y))$

(BI1) $\forall x.\forall Q_{1},Q_{2}\in D_{a}.(Q_{1}\in R_{a}^{I}(x)\wedge
\{y$ $|$ $(x,y)\in R_{a}^{B}\}\subseteq (Q_{1}\cap Q_{2})\cup (\overline{%
Q_{1}}\cap \overline{Q_{2}}))\rightarrow Q_{2}\in R_{a}^{I}(x)$

(BI2) $\forall x.\forall Q.Q\in R_{a}^{I}(x)\rightarrow (\forall y.(x,y)\in
R_{a}^{B}\rightarrow Q\in R_{a}^{I}(y))$

(BI3) $\forall x.\forall Q\in D_{a}.(\exists y.(x,y)\in R_{a}^{B}\wedge Q\in
R_{a}^{I}(y))\rightarrow Q\in R_{a}^{I}(x)$

(BI4) $\forall x.\forall Q\in D_{a}.(\forall y.(x,y)\in R_{a}^{B}\rightarrow
Q\in R_{a}^{I}(y))\rightarrow Q\in R_{a}^{I}(x)$

(BI5) $\forall x.\forall Q\in D_{a}.Q\in R_{a}^{I}(x)\rightarrow \exists
y.((x,y)\in R_{a}^{B}\wedge Q\in R_{a}^{I}(y))$

(BPIEF1a) $\forall x.\forall Q\in D_{a}.(Q\in R_{a}^{I}(x)\rightarrow Q\in
R_{a}^{P}(x))$

(BPIEF1b) $\forall x.(\cup \{Q$ \TEXTsymbol{\vert} $Q\in R_{a}^{I}(x))\}\cap
\{y$ \TEXTsymbol{\vert} $(x,y)\in R_{a}^{B}\}=\varnothing $

(BPIEF1c) $\forall x.\forall y.\forall Q\in D_{a}.((x,y)\in R_{a}^{B}\wedge
Q\in R_{a}^{I}(x))\rightarrow \exists z.((y,z)\in R_{a}^{EF}\wedge z\in Q),$
where $R_{a}^{EF}$ is the reflexive and transitive closure of $R_{a}^{X}.$

(BX1) $\forall x.\{y$ $|$ $(x,y)\in R_{a}^{B}\circ R_{a}^{X}\circ
R_{a}^{B}\}\subseteq \{y$ $|$ $(x,y)\in R_{a}^{B}\circ R_{a}^{X}\}$

(BX2) $\forall x.\forall Q\in D_{a}.(\forall y.\exists z.(x,y)\in
R_{a}^{B}\rightarrow ((y,z)\in R_{a}^{X}\wedge z\in Q))\rightarrow (\forall
u.\exists v.\forall w.(x,u)\in R_{a}^{B}\rightarrow (((u,v)\in
R_{a}^{X})\wedge ((v,w)\in R_{a}^{B}\rightarrow w\in Q)))$

A simple example model of $BPICTL$ is $M=(S,$ $\Pi ,$ $\pi ,$ $%
R_{a}^{B}=S\times S,$ $R_{a}^{P}(s)=\{S\}$ for any $s,$ $R_{a}^{I}(s)=%
\varnothing $ for any $s,$ $R_{a}^{X}=\{(s,s)$ \TEXTsymbol{\vert} $s\in S\},$
$D_{a}=\wp (S)),$ where $S\neq \varnothing ,$ $\Pi \neq \varnothing $ and $%
\pi $ is an arbitrary map: $S\times \Pi \longrightarrow \{true,false\}.$

Formally, a formula $\varphi $ is interpreted as a set of states in which $%
\varphi $ is true. We write such set of states as $[[\varphi ]]_{M},$ where $%
M$ is a model$.$ The set $[[\varphi ]]_{M}$ is defined recursively as
follows:

\textbf{Definition 3.} Semantics of $BPICTL$

$[[p]]_{M}=\{s$ $|$ $\pi (s,p)=true\}$

$[[\lnot \varphi ]]_{M}=S-[[\varphi ]]_{M}$

$[[\varphi _{1}\wedge \varphi _{2}]]_{M}=[[\varphi _{1}]]_{M}\cap \lbrack
\lbrack \varphi _{2}]]_{M}$

$[[\varphi _{1}\vee \varphi _{2}]]_{M}=[[\varphi _{1}]]_{M}\cup \lbrack
\lbrack \varphi _{2}]]_{M}$

$[[B_{a}\varphi ]]_{M}=\{s$ $|$ for all $(s,t)\in R_{a}^{B}$ implies $t\in
\lbrack \lbrack \varphi ]]_{M}.\}$

$[[P_{a}\varphi ]]_{M}=\{s$ $|$ $[[\varphi ]]_{M}\in R_{a}^{P}(s).\}$

$[[I_{a}\varphi ]]_{M}=\{s$ $|$ $[[\varphi ]]_{M}\in R_{a}^{I}(s).\}$

$[[AX\varphi ]]_{M}=\{s$ $|$ for all $(s,t)\in R_{a}^{X}$ implies $t\in
\lbrack \lbrack \varphi ]]_{M}.\}$

$[[EX\varphi ]]_{M}=\{s$ $|$ there exists $(s,t)\in R_{a}^{X}$ such that $%
t\in \lbrack \lbrack \varphi ]]_{M}.\}$

$[[EF\varphi ]]_{M}=\{s$ $|$ there exists infinite trace $\rho ,$ $\rho
=t_{0},t_{1},...$ $,$ where $t_{0}=s$ and $\forall
t_{i},t_{i+1}.(t_{i},t_{i+1})\in R_{a}^{X}\wedge \exists t_{m}.t_{m}\in
\lbrack \lbrack \varphi ]]_{M}.\}$

$[[EG\varphi ]]_{M}=\{s$ $|$ there exists infinite trace $\rho ,$ $\rho
=t_{0},t_{1},...$ $,$ where $t_{0}=s$ and $\forall
t_{i},t_{i+1}.(t_{i},t_{i+1})\in R_{a}^{X}\wedge \forall t_{i}.t_{i}\in
\lbrack \lbrack \varphi ]]_{M}.\}$

$[[E\varphi _{1}U\varphi _{2}]]_{M}=\{s$ $|$ there exists infinite trace $%
\rho =t_{0},t_{1},...$ $,$ where $t_{0}=s$, $\forall
t_{i},t_{i+1}.(t_{i},t_{i+1})\in R_{a}^{X},$ and there exists a position $%
m\geq 0,$ such that $t_{m}\in \lbrack \lbrack \varphi _{2}]]_{M}$ and for
all positions $0\leq k<m,$ we have $t_{k}\in \lbrack \lbrack \varphi
_{1}]]_{M}.\}$

In order to characterize the properties of belief, preference, intention and
temporal, we will characterize the formulas that are always true. More
formally, given a model $M$, we say that $\varphi $ is valid in $M$, and
write $M\models \varphi $, if $s\in \lbrack \lbrack \varphi ]]_{M}$ for
every state $s$ in $S$, and we say that $\varphi $ is satisfiable in $M$,
and write $(M,s)\models \varphi $, if $s\in \lbrack \lbrack \varphi ]]_{M}$
for some $s$ in $S$. We say that $\varphi $ is valid, and write $\models
_{BPICTL}\varphi $, if $\varphi $ is valid in all models, and that $\varphi $
is satisfiable if it is satisfiable in some model. We write $\Gamma \models
_{BPICTL}\varphi $, if $\varphi $ is valid in all models in which $\Gamma $
is satisfiable.

\subsection{Inference System of $BPICTL$}

Now we list a number of valid properties of belief, preference, intention
and temporal, which form the inference system of $BPICTL$.

\textit{All} \textit{instances} \textit{of} \textit{propositional} \textit{%
tautologies} \textit{and} \textit{rules}.

(B1) $\vdash \varphi \Rightarrow \vdash B_{a}\varphi $

(B2) $(B_{a}\varphi _{1}\wedge B_{a}(\varphi _{1}\rightarrow \varphi
_{2}))\rightarrow B_{a}\varphi _{2}$

(B3) $B_{a}\varphi \rightarrow B_{a}B_{a}\varphi $

(B4) $\lnot B_{a}\varphi \rightarrow B_{a}\lnot B_{a}\varphi $

(B5) $B_{a}\varphi \rightarrow \lnot B_{a}\lnot \varphi $

(P1) $(P_{a}\varphi _{1}\wedge P_{a}\varphi _{2})\rightarrow P_{a}(\varphi
_{1}\wedge \varphi _{2})$

(P2) $(P_{a}\varphi _{1}\wedge P_{a}(\varphi _{1}\rightarrow \varphi
_{2}))\rightarrow P_{a}\varphi _{2}$

(P3) $P_{a}P_{a}\varphi \rightarrow P_{a}\varphi $

(P4) $P_{a}\lnot P_{a}\varphi \rightarrow \lnot P_{a}\varphi $

(AX1) $\vdash \varphi \Rightarrow \vdash AX\varphi $

(AX2) $(AX\varphi _{1}\wedge AX(\varphi _{1}\rightarrow \varphi
_{2}))\rightarrow AX\varphi _{2}$

(AX3) $AX\varphi \rightarrow \lnot AX\lnot \varphi $

(EX1) $EX\varphi \leftrightarrow \lnot AX\lnot \varphi $

(EF1) $EF\varphi \leftrightarrow EtrueU\varphi $

(EG1) $EG\varphi \leftrightarrow (\varphi \wedge EXEG\varphi )$

(EG2) $(\psi \leftrightarrow (\varphi \wedge EX\psi ))\rightarrow (\psi
\rightarrow EG\varphi )$

(EU1) $E\varphi _{1}U\varphi _{2}\leftrightarrow (\varphi _{2}\vee (\varphi
_{1}\wedge EXE\varphi _{1}U\varphi _{2}))$

(EU2) $(\psi \leftrightarrow (\varphi _{2}\vee (\varphi _{1}\wedge EX\psi
)))\rightarrow (E\varphi _{1}U\varphi _{2}\rightarrow \psi )$

(BP1) $(B_{a}(\varphi _{1}\leftrightarrow \varphi _{2})\wedge P_{a}\varphi
_{1})\rightarrow P_{a}\varphi _{2}.$

(BP2) $P_{a}\varphi \rightarrow B_{a}P_{a}\varphi $

(BP3) $\lnot P_{a}\varphi \rightarrow B_{a}\lnot P_{a}\varphi $

(BP4) $B_{a}P_{a}\varphi \rightarrow P_{a}\varphi $

(BP5) $B_{a}\lnot P_{a}\varphi \rightarrow \lnot P_{a}\varphi $

(BI1) $(B_{a}(\varphi _{1}\leftrightarrow \varphi _{2})\wedge I_{a}\varphi
_{1})\rightarrow I_{a}\varphi _{2}$

(BI2) $I_{a}\varphi \rightarrow B_{a}I_{a}\varphi $

(BI3) $\lnot I_{a}\varphi \rightarrow B_{a}\lnot I_{a}\varphi $

(BI4) $B_{a}I_{a}\varphi \rightarrow I_{a}\varphi $

(BI5) $B_{a}\lnot I_{a}\varphi \rightarrow \lnot I_{a}\varphi $

(BPIEF1) $I_{a}\varphi \rightarrow (P_{a}\varphi \wedge B_{a}\lnot \varphi
\wedge B_{a}EF\varphi )$

(BX1) $B_{a}AX\varphi \rightarrow B_{a}AXB_{a}\varphi $

(BX2) $B_{a}EX\varphi \rightarrow B_{a}EXB_{a}\varphi $

In this inference system, B1-B5 characterize belief modality. P1-P4
characterize preference modality. AX1, AX2, AX3, EX1, EF1, EG1, EG2, EU1,
EU2 characterize temporal operators, BP1-BP5 characterize the interaction of
belief and preference, BI1-BI5 characterize the interaction of belief and
intention, BPIEF1 characterizes the interaction of belief, preference,
intention and temporal operators, BX1, BX2 characterize the interaction of
belief and temporal operators. Although some interaction properties between
mental modal operators were studied in previous works, interactive
properties between mental and temporal modal operators are rarely studied.
In the inference system of $BPICTL$, we present some new interactive
properties between belief modality, preference modality, intention modality
and temporal operators. For example, $I_{a}\varphi \rightarrow (P_{a}\varphi
\wedge B_{a}\lnot \varphi \wedge B_{a}EF\varphi )$ means that agent $a$
intends $\varphi $ implies that agent $a$ prefers $\varphi $ and agent $a$
believes that $\varphi $ is not true now but may be true in the future.

A proof in $BPICTL$ consists of a sequence of formulas, each of which is
either an instance of an axiom in $BPICTL$ or follows from an application of
an inference rule. (If \textquotedblleft $\varphi _{1},...,\varphi _{n}$
infer $\psi $\textquotedblright\ is an instance of an inference rule, and if
the formulas $\varphi _{1},...,\varphi _{n}$ have appeared earlier in the
proof, then we say that $\psi $ follows from an application of an inference
rule.) A proof is said to be from $\Gamma $ to $\varphi $ if the premise is $%
\Gamma $ and the last formula is $\varphi $ in the proof. We say $\varphi $
is provable from $\Gamma $ in $BPICTL$, and write $\Gamma \vdash
_{BPICTL}\varphi $, if there is a proof from $\Gamma $ to $\varphi $ in $%
BPICTL$.

\section{Soundness and Completeness of $BPICTL$}

Inference system of $BPICTL$ is said to be sound with respect to $BPICTL$
models if every formula provable in $BPICTL$ is valid with respect to
models. The system $BPICTL$ is complete with respect to models if every
formula valid with respect to models is provable in $BPICTL$. The soundness
and completeness provide a tight connection between the syntactic notion of
provability and the semantic notion of validity.

In the following, we prove that all axioms and rules in the inference system
hold in any model $M.$ Therefore we have the soundness of the inference
system:

\textbf{Proposition 1} (Soundness of $BPICTL$). The inference system of $%
BPICTL$ is sound, i.e., $\Gamma \vdash _{BPICTL}\varphi \Rightarrow \Gamma
\models _{BPICTL}\varphi .$

$Proof:$ We show each axiom and each rule of $BPICTL$ is sound, respectively.

(B1) $\vdash \varphi \Rightarrow \vdash B_{a}\varphi :$ It is trivial by the
soundness of belief logic.

(B2) $(B_{a}\varphi \wedge B_{a}(\varphi \rightarrow \psi ))\rightarrow
B_{a}\psi :$ It is trivial by the soundness of belief logic.

(B3) $B_{a}\varphi \rightarrow B_{a}B_{a}\varphi :$ It is trivial by the
soundness of belief logic.

(B4) $\lnot B_{a}\varphi \rightarrow B_{a}\lnot B_{a}\varphi :$ It is
trivial by the soundness of belief logic.

(B5) $B_{a}\varphi \rightarrow \lnot B_{a}\lnot \varphi :$ It is trivial by
the soundness of belief logic.

(P1) $(P_{a}\varphi _{1}\wedge P_{a}\varphi _{2})\rightarrow P_{a}(\varphi
_{1}\wedge \varphi _{2}):$ By $\forall x.\forall Q_{1},Q_{2}.(Q_{1}\in
R_{a}^{P}(x)\wedge Q_{2}\in R_{a}^{P}(x)\rightarrow Q_{1}\cap Q_{2}\in
R_{a}^{P}(x)).$

(P2) $(P_{a}\varphi _{1}\wedge P_{a}(\varphi _{1}\rightarrow \varphi
_{2}))\rightarrow P_{a}\varphi _{2}:$ By $\forall x.\forall
Q_{1},Q_{2}.(Q_{1}\in R_{a}^{P}(x)\wedge (\overline{Q_{1}}\cup Q_{2})\in
R_{a}^{P}(x)\rightarrow Q_{2}\in R_{a}^{P}(x)).$

(P3) $P_{a}P_{a}\varphi \rightarrow P_{a}\varphi :$ Suppose $x\in \lbrack
\lbrack P_{a}P_{a}\varphi ]]_{M},$ we have $[[P_{a}\varphi ]]_{M}\in
R_{a}^{P}(x).$ Then $\{y$ $|$ $[[\varphi ]]_{M}\in R_{a}^{P}(y)\}\in
R_{a}^{P}(x).$ Since $\forall x.\forall Q.(\{y$ $|$ $Q\in R_{a}^{P}(y)\}\in
R_{a}^{P}(x))\rightarrow Q\in R_{a}^{P}(x),$ we have $[[\varphi ]]_{M}\in
R_{a}^{P}(x),$ $x\in \lbrack \lbrack P_{a}\varphi ]]_{M}.$ Therefore $%
P_{a}P_{a}\varphi \rightarrow P_{a}\varphi .$

(P4) $P_{a}\lnot P_{a}\varphi \rightarrow \lnot P_{a}\varphi :$ Suppose $%
x\in \lbrack \lbrack P_{a}\varphi ]]_{M},$ we have $[[\varphi ]]_{M}\in
R_{a}^{P}(x).$ Since $\forall x.\forall Q_{1}.Q_{1}\in
R_{a}^{P}(x)\rightarrow \forall Q_{2}.\exists y.((Q_{2}\in
R_{a}^{P}(x)\rightarrow (y\in Q_{2}\wedge Q_{1}\in R_{a}^{P}(y))),$ $\forall
Q_{2}.\exists y.((Q_{2}\in R_{a}^{P}(x)\rightarrow (y\in Q_{2}\wedge \lbrack
\lbrack \varphi ]]_{M}\in R_{a}^{P}(y))).$ Therefore $x\in \lbrack \lbrack
\lnot P_{a}\lnot P_{a}\varphi ]]_{M}.$ We have $P_{a}\varphi \rightarrow
\lnot P_{a}\lnot P_{a}\varphi .$

(AX1) $\vdash \varphi \Rightarrow \vdash AX\varphi :$ It is trivial by the
soundness of temporal logic $CTL$.

(AX2) $(AX\varphi _{1}\wedge AX(\varphi _{1}\rightarrow \varphi
_{2}))\rightarrow AX\varphi _{2}:$ It is trivial by the soundness of
temporal logic $CTL$.

(AX3) $AX\varphi \rightarrow \lnot AX\lnot \varphi :$ It is trivial by the
soundness of temporal logic $CTL$.

(EX1) $EX\varphi \leftrightarrow \lnot AX\lnot \varphi :$ It is trivial by
the soundness of temporal logic $CTL$.

(EF1) $EF\varphi \leftrightarrow EtrueU\varphi :$ It is trivial by the
soundness of temporal logic $CTL$.

(EG1) $EG\varphi \leftrightarrow (\varphi \wedge EXEG\varphi ):$ It is
trivial by the soundness of temporal logic $CTL$.

(EG2) $(\psi \leftrightarrow (\varphi \wedge EX\psi ))\rightarrow (\psi
\rightarrow EG\varphi ):$ It is trivial by the soundness of temporal logic $%
CTL$.

(EU1) $E\varphi _{1}U\varphi _{2}\leftrightarrow (\varphi _{2}\vee (\varphi
_{1}\wedge EXE\varphi _{1}U\varphi _{2})):$ It is trivial by the soundness
of temporal logic $CTL$.

(EU2) $(\psi \leftrightarrow (\varphi _{2}\vee (\varphi _{1}\wedge EX\psi
)))\rightarrow (E\varphi _{1}U\varphi _{2}\rightarrow \psi ):$ It is trivial
by the soundness of temporal logic $CTL$.

(BP1) $(B_{a}(\varphi _{1}\leftrightarrow \varphi _{2})\wedge P_{a}\varphi
_{1})\rightarrow P_{a}\varphi _{2}.:$ Suppose $x\in \lbrack \lbrack
B_{a}(\varphi _{1}\leftrightarrow \varphi _{2})]]_{M},$ $x\in \lbrack
\lbrack P_{a}\varphi _{1}]]_{M},$ we have $[[\varphi _{1}]]_{M}\in
R_{a}^{P}(x)\wedge \{y$ $|$ $(x,y)\in R_{a}^{B}\}\subseteq ([[\varphi
_{1}]]_{M}\cap \lbrack \lbrack \varphi _{2}]]_{M})\cup (\overline{[[\varphi
_{1}]]_{M}}\cap \overline{[[\varphi _{2}]]_{M}}).$ Since $\forall x.\forall
Q_{1},Q_{2}\in D_{a}.(Q_{1}\in R_{a}^{P}(x)\wedge \{y$ $|$ $(x,y)\in
R_{a}^{B}\}\subseteq (Q_{1}\cap Q_{2})\cup (\overline{Q_{1}}\cap \overline{%
Q_{2}}))\rightarrow Q_{2}\in R_{a}^{P}(x),$ $x\in \lbrack \lbrack
P_{a}\varphi _{2}]]_{M}.$ We have $(B_{a}(\varphi _{1}\leftrightarrow
\varphi _{2})\wedge P_{a}\varphi _{1})\rightarrow P_{a}\varphi _{2}.$

(BP2) $P_{a}\varphi \rightarrow B_{a}P_{a}\varphi :$ Suppose $x\in \lbrack
\lbrack P_{a}\varphi ]]_{M},$ we have $[[\varphi ]]_{M}\in R_{a}^{P}(x).$ By 
$\forall x.\forall Q.Q\in R_{a}^{P}(x)\rightarrow (\forall y.(x,y)\in
R_{a}^{B}\rightarrow Q\in R_{a}^{P}(y)),$ we have $\forall y.((x,y)\in
R_{a}^{B}\rightarrow \lbrack \lbrack \varphi ]]_{M}\in R_{a}^{P}(y)).$ So $%
x\in \lbrack \lbrack B_{a}P_{a}\varphi ]]_{M}.$ We have $P_{a}\varphi
\rightarrow B_{a}P_{a}\varphi .$

(BP3) $\lnot P_{a}\varphi \rightarrow B_{a}\lnot P_{a}\varphi :$ Suppose $%
x\in \lbrack \lbrack \lnot B_{a}\lnot P_{a}\varphi ]]_{M},$ we have $\exists
y.(x,y)\in R_{a}^{B}\wedge \lbrack \lbrack \varphi ]]_{M}\in R_{a}^{P}(y).$
Since $\forall x.\forall Q.(\exists y.(x,y)\in R_{a}^{B}\wedge Q\in
R_{a}^{P}(y))\rightarrow Q\in R_{a}^{P}(x),$ then $[[\varphi ]]_{M}\in
R_{a}^{P}(x).$ Therefore $x\in \lbrack \lbrack P_{a}\varphi ]]_{M}.$ We have 
$\lnot B_{a}\lnot P_{a}\varphi \rightarrow P_{a}\varphi ,$ so $\lnot
P_{a}\varphi \rightarrow B_{a}\lnot P_{a}\varphi .$

(BP4) $B_{a}P_{a}\varphi \rightarrow P_{a}\varphi :$ Suppose $x\in \lbrack
\lbrack B_{a}P_{a}\varphi ]]_{M},$ we have $\forall x.(\forall y.(x,y)\in
R_{a}^{B}\rightarrow \lbrack \lbrack \varphi ]]_{M}\in R_{a}^{P}(y)).$ Since 
$\forall x.\forall Q.(\forall y.(x,y)\in R_{a}^{B}\rightarrow Q\in
R_{a}^{P}(y))\rightarrow Q\in R_{a}^{P}(x),$ we have $[[\varphi ]]_{M}\in
R_{a}^{P}(x),$ $x\in \lbrack \lbrack P_{a}\varphi ]]_{M}.$ Therefore $%
B_{a}P_{a}\varphi \rightarrow P_{a}\varphi .$

(BP5) $B_{a}\lnot P_{a}\varphi \rightarrow \lnot P_{a}\varphi :$ Suppose $%
[[\varphi ]]_{M}\in R_{a}^{P}(x),$ we have $x\in \lbrack \lbrack
P_{a}\varphi ]]_{M}.$ Since $\forall x.\forall Q.Q\in
R_{a}^{P}(x)\rightarrow \exists y.((x,y)\in R_{a}^{B}\wedge Q\in
R_{a}^{P}(y)),$ $\exists y.(x,y)\in R_{a}^{B}\wedge \lbrack \lbrack \varphi
]]_{M}\in R_{a}^{P}(y),$ So $x\in \lbrack \lbrack \lnot B_{a}\lnot
P_{a}\varphi ]]_{M}.$ Therefore $P_{a}\varphi \rightarrow \lnot B_{a}\lnot
P_{a}\varphi .$

(BI1) $(B_{a}(\varphi _{1}\leftrightarrow \varphi _{2})\wedge I_{a}\varphi
_{1})\rightarrow I_{a}\varphi _{2}:$ Suppose $x\in \lbrack \lbrack
B_{a}(\varphi _{1}\leftrightarrow \varphi _{2})]]_{M},$ $x\in \lbrack
\lbrack I_{a}\varphi _{1}]]_{M}.$ Since $\forall x.\forall Q_{1},Q_{2}\in
D_{a}.(Q_{1}\in R_{a}^{I}(x)\wedge \{y$ $|$ $(x,y)\in R_{a}^{B}\}\subseteq
(Q_{1}\cap Q_{2})\cup (\overline{Q_{1}}\cap \overline{Q_{2}}))\rightarrow
Q_{2}\in R_{a}^{I}(x),$ $x\in \lbrack \lbrack I_{a}\varphi _{2}]]_{M}.$ We
have $(B_{a}(\varphi _{1}\leftrightarrow \varphi _{2})\wedge I_{a}\varphi
_{1})\rightarrow I_{a}\varphi _{2}.$

(BI2) $I_{a}\varphi \rightarrow B_{a}I_{a}\varphi :$ Suppose $x\in \lbrack
\lbrack I_{a}\varphi ]]_{M},$ we have $[[\varphi ]]_{M}\in R_{a}^{I}(x).$ By 
$\forall x.\forall Q.Q\in R_{a}^{I}(x)\rightarrow (\forall y.(x,y)\in
R_{a}^{B}\rightarrow Q\in R_{a}^{I}(y)),$ we have $\forall y.(x,y)\in
R_{a}^{B}\rightarrow \lbrack \lbrack \varphi ]]_{M}\in R_{a}^{I}(y)$. So $%
x\in \lbrack \lbrack B_{a}I_{a}\varphi ]]_{M}.$ We have $I_{a}\varphi
\rightarrow B_{a}I_{a}\varphi .$

(BI3) $\lnot I_{a}\varphi \rightarrow B_{a}\lnot I_{a}\varphi :$ Suppose $%
x\in \lbrack \lbrack \lnot B_{a}\lnot I_{a}\varphi ]]_{M},$ we have $\exists
y.(x,y)\in R_{a}^{B}\wedge \lbrack \lbrack \varphi ]]_{M}\in R_{a}^{I}(y).$
Since $\forall x.\forall Q.(\exists y.(x,y)\in R_{a}^{B}\wedge Q\in
R_{a}^{I}(y))\rightarrow Q\in R_{a}^{I}(x),$ then $[[\varphi ]]_{M}\in
R_{a}^{I}(x).$ Therefore $x\in \lbrack \lbrack I_{a}\varphi ]]_{M}.$ We have 
$\lnot B_{a}\lnot I_{a}\varphi \rightarrow I_{a}\varphi ,$ so $\lnot
I_{a}\varphi \rightarrow B_{a}\lnot I_{a}\varphi .$

(BI4) $B_{a}I_{a}\varphi \rightarrow I_{a}\varphi :$ Suppose $x\in \lbrack
\lbrack B_{a}I_{a}\varphi ]]_{M},$ we have $\forall y.(x,y)\in
R_{a}^{B}\rightarrow y\in \lbrack \lbrack I_{a}\varphi ]]_{M}.$ Then $%
\forall x.(\forall y.(x,y)\in R_{a}^{B}\rightarrow \lbrack \lbrack \varphi
]]_{M}\in R_{a}^{I}(y)).$ Since $\forall x.\forall Q.(\forall y.(x,y)\in
R_{a}^{B}\rightarrow Q\in R_{a}^{I}(y))\rightarrow Q\in R_{a}^{I}(x),$ we
have $[[\varphi ]]_{M}\in R_{a}^{I}(x),$ $x\in \lbrack \lbrack I_{a}\varphi
]]_{M}.$ Therefore $B_{a}I_{a}\varphi \rightarrow I_{a}\varphi .$

(BI5) $B_{a}\lnot I_{a}\varphi \rightarrow \lnot I_{a}\varphi :$ Suppose $%
[[\varphi ]]_{M}\in R_{a}^{I}(x),$ we have $x\in \lbrack \lbrack
I_{a}\varphi ]]_{M}.$ Since $\forall x.\forall Q.Q\in
R_{a}^{I}(x)\rightarrow \exists y.((x,y)\in R_{a}^{B}\wedge Q\in
R_{a}^{I}(y)),$ $\exists y.(x,y)\in R_{a}^{B}\wedge \lbrack \lbrack \varphi
]]_{M}\in R_{a}^{I}(y).$ So $x\in \lbrack \lbrack \lnot B_{a}\lnot
I_{a}\varphi ]]_{M}.$ Therefore $I_{a}\varphi \rightarrow \lnot B_{a}\lnot
I_{a}\varphi .$

(BPIEF1) $I_{a}\varphi \rightarrow (P_{a}\varphi \wedge B_{a}\lnot \varphi
\wedge B_{a}EF\varphi ):$ \ It is enough to prove that $(I_{a}\varphi
\rightarrow P_{a}\varphi )\wedge (I_{a}\varphi \rightarrow B_{a}\lnot
\varphi )\wedge (I_{a}\varphi \rightarrow B_{a}EF\varphi )$.

(BPIEF1a) $I_{a}\varphi \rightarrow P_{a}\varphi :$ Suppose $x\in \lbrack
\lbrack I_{a}\varphi ]]_{M},$ then $[[\varphi ]]_{M}\in R_{a}^{I}(x).$ By $%
\forall x.\forall Q.(Q\in R_{a}^{I}(x)\rightarrow Q\in R_{a}^{P}(x)),$ we
have $[[\varphi ]]_{M}\in R_{a}^{P}(x).$ Hence $x\in \lbrack \lbrack
P_{a}\varphi ]]_{M}.$

(BPIEF1b) $I_{a}\varphi \rightarrow B_{a}\lnot \varphi :$ Suppose $x\in
\lbrack \lbrack I_{a}\varphi ]]_{M},$ then $[[\varphi ]]_{M}\in
R_{a}^{I}(x). $ Since $\forall x.(\cup \{Q$ \TEXTsymbol{\vert} $Q\in
R_{a}^{I}(x))\}\cap \{y$ \TEXTsymbol{\vert} $(x,y)\in
R_{a}^{B}\}=\varnothing ,$ we have $[[\varphi ]]_{M}\cap \{y$ \TEXTsymbol{%
\vert} $(x,y)\in R_{a}^{B}\}=\varnothing ,$ $\{y$ \TEXTsymbol{\vert} $%
(x,y)\in R_{a}^{B}\}\subseteq \lbrack \lbrack \lnot \varphi ]]_{M}.$ So $%
x\in \lbrack \lbrack B_{a}\lnot \varphi ]]_{M}.$

(BPIEF1c) $I_{a}\varphi \rightarrow B_{a}EF\varphi :$ Suppose $x\in \lbrack
\lbrack I_{a}\varphi ]]_{M},$ then $[[\varphi ]]_{M}\in R_{a}^{I}(x).$ By $%
\forall x.\forall y.\forall Q\in D_{a}.((x,y)\in R_{a}^{B}\wedge Q\in
R_{a}^{I}(x))\rightarrow \exists z.((y,z)\in R_{a}^{EF}\wedge z\in Q),$ we
have $\forall y.$ $(x,y)\in R_{a}^{B}\rightarrow \exists z.((y,z)\in
R_{a}^{EF}\wedge z\in \lbrack \lbrack \varphi ]]_{M}).$ Hence $x\in \lbrack
\lbrack B_{a}EF\varphi ]]_{M}.$

(BX1) $B_{a}AX\varphi \rightarrow B_{a}AXB_{a}\varphi :$ Suppose $x\in
\lbrack \lbrack B_{a}AX\varphi ]]_{M},$ then $\forall y.(x,y)\in
R_{a}^{B}\circ R_{a}^{X}\rightarrow y\in \lbrack \lbrack \varphi ]]_{M}$
where $R_{1}\circ R_{2}=\{(x,z)$ $|$ $\exists y.(x,y)\in R_{1}\wedge
(y,z)\in R_{2}\}.$ By $\forall x.\{y$ $|$ $(x,y)\in R_{a}^{B}\circ
R_{a}^{X}\circ R_{a}^{B}\}\subseteq \{y$ $|$ $(x,y)\in R_{a}^{B}\circ
R_{a}^{X}\},$ we have $\{y$ $|$ $(x,y)\in R_{a}^{B}\circ R_{a}^{X}\circ
R_{a}^{B}\}\subseteq \lbrack \lbrack \varphi ]]_{M}.$ Hence $x\in \lbrack
\lbrack B_{a}AXB_{a}\varphi ]]_{M}.$

(BX2) $B_{a}EX\varphi \rightarrow B_{a}EXB_{a}\varphi :$ Suppose $x\in
\lbrack \lbrack B_{a}EX\varphi ]]_{M},$ then $\forall y.\exists z.(x,y)\in
R_{a}^{B}\rightarrow ((y,z)\in R_{a}^{X}\wedge z\in \lbrack \lbrack \varphi
]]_{M}).$ By $\forall x.\forall Q\in D_{a}.(\forall y.\exists z.(x,y)\in
R_{a}^{B}\rightarrow ((y,z)\in R_{a}^{X}\wedge z\in Q))\rightarrow (\forall
u.\exists v.\forall w.(x,u)\in R_{a}^{B}\rightarrow ((u,v)\in
R_{a}^{X})\wedge ((v,w)\in R_{a}^{B}\rightarrow w\in Q)),$ we have $\forall
y.\exists z.\forall w.(x,y)\in R_{a}^{B}\rightarrow ((y,z)\in
R_{a}^{X})\wedge ((z,w)\in R_{a}^{B}\rightarrow w\in \lbrack \lbrack \varphi
]]_{M}))$. Hence $x\in \lbrack \lbrack B_{a}EXB_{a}\varphi ]]_{M}.$%
%TCIMACRO{\TeXButton{End Proof}{\endproof}}%
%BeginExpansion
\endproof%
%EndExpansion

We shall show that the inference system of $BPICTL$ provides a complete
axiomatization for belief, preference, intention and temporal property with
respect to a $BPICTL$ model. To achieve this aim, it suffices to prove that
every $BPICTL$-consistent set is satisfiable with respect to a $BPICTL$
model. We construct a special structure $M$ called a canonical structure for 
$BPICTL$. $M$ has a state $s_{V}$ corresponding to every maximal $BPICTL$%
-consistent set $V$ and the following property holds: $(M,s_{V})\models
\varphi $ iff $\varphi \in V$.

We need some definitions before giving the proof of the completeness. Given
an inference system of $BPICTL$, we say a set of formulae $\Gamma $ is a
consistent set with respect to $L^{BPICTL}$ exactly if false is not provable
from $\Gamma $. A set of formulae $\Gamma $ is a maximal consistent set with
respect to $L^{BPICTL}$ if (1) it is $BPICTL$-consistent, and (2) for all $%
\varphi $ in $L^{BPICTL}$ but not in $\Gamma $, the set $\Gamma \cup
\{\varphi \}$ is not $BPICTL$-consistent.

\textbf{Definition 4.} The canonical model $M$ with respect to $BPICTL$ is $%
(S,$ $\Pi ,$ $\pi ,$ $R_{a}^{B},$ $R_{a}^{P},$ $R_{a}^{I},$ $R_{a}^{X},$ $%
D_{a})$.

(1) $S=\{\Gamma $ $|$ $\Gamma $ is a maximal consistent set with respect to $%
BPICTL\}$.

(2) $\Pi $ is the set of atomic formulae.

(3) $\pi $ is a truth assignment as follows: for any atomic formula $p$, $%
\pi (\Gamma ,p)=true\Leftrightarrow p\in \Gamma $.

(4) $R_{a}^{B}=\{(\Gamma _{1},\Gamma _{2})$ $|$ $\Gamma _{1}/B_{a}\subseteq
\Gamma _{2}\},$where $\Gamma _{1}/B_{a}=\{\varphi $ $|$ $B_{a}\varphi \in
\Gamma _{1}\}.$

(5) $R_{a}^{P}$ maps every element of $S$ to a subset of $\wp (S)$: $%
R_{a}^{P}(\Gamma )=\{U_{a}^{P}(\varphi )$ $|$ $\varphi $ is a formula of $%
BPICTL$ and $P_{a}\varphi \in \Gamma ,$where $U_{a}^{P}(\varphi )=\{\Gamma
^{\prime }$ $|$ $\varphi \in \Gamma ^{\prime }\ $and $P_{a}\varphi \in
\Gamma \}\}$.

(6) $R_{a}^{I}$ maps every element of $S$ to a subset of $\wp (S)$: $%
R_{a}^{I}(\Gamma )=\{U_{a}^{I}(\varphi )$ $|$ $\varphi $ is a formula of $%
BPICTL$ and $I_{a}\varphi \in \Gamma ,$where $U_{a}^{I}(\varphi )=\{\Gamma
^{\prime }$ $|$ $\varphi \in \Gamma ^{\prime }\ $and $I_{a}\varphi \in
\Gamma \}\}$.

(7) $R_{a}^{X}=\{(\Gamma _{1},\Gamma _{2})$ $|$ $\Gamma _{1}/AX_{a}\subseteq
\Gamma _{2}\},$where $\Gamma _{1}/AX_{a}=\{\varphi $ $|$ $AX\varphi \in
\Gamma _{1}\}.$

(8) $D_{a}=\{\{\Gamma _{\varphi }$ $|$ $\varphi \in \Gamma _{\varphi }\}$ 
\TEXTsymbol{\vert} $\varphi $ is an arbitrary formula of $BPICTL\}.$

\textbf{Lemma 1.} $S$ is a nonempty set.

$Proof.$ Since the rules and axioms of $BPICTL$ are consistent, $S$ is
nonempty.%
%TCIMACRO{\TeXButton{End Proof}{\endproof}}%
%BeginExpansion
\endproof%
%EndExpansion

In classical logic, it is easy to see that every consistent set of formulae
can be extended to a maximal consistent set. Therefore we have the following
lemma.

\textbf{Lemma 2.} For any $BPICTL$-consistent set of formulae $\Delta $,
there is a maximal $BPICTL$-consistent set $\Gamma $ such that $\Delta
\subseteq \Gamma $.

$Proof.$ To show that $\Delta $ can be extended to a maximal $BPICTL$%
-consistent set, we construct a sequence $\Gamma _{0}$, $\Gamma _{1}$, ...
of $BPICTL$-consistent sets as follows.

Let $\psi _{1}$, $\psi _{2}$, ... be a sequence of the formulae in $%
L^{BPICTL}$.

At first, we construct $\Gamma _{i}$ which satisfies the following
conditions:

(1) $\Delta =\Gamma _{0}\subseteq \Gamma _{i}$.

(2) $\Gamma _{i}$ is consistent.

(3) For every $\psi _{i+1}\in L^{BPICTL}$, either $\{\psi _{i+1}\}\cup
\Gamma _{i}$ or $\{\lnot \psi _{i+1}\}\cup \Gamma _{i}$ is consistent. We
let $\Gamma _{i+1}=\{\psi _{i+1}\}\cup \Gamma _{i}$ if $\{\psi _{i+1}\}\cup
\Gamma _{i}$ is consistent. $\Gamma _{i+1}=\{\lnot \psi _{i+1}\}\cup \Gamma
_{i}$ if $\{\lnot \psi _{i+1}\}\cup \Gamma _{i}$ is consistent.

We\ let $\Gamma =\cup _{i\geq 0}\Gamma _{i}.$ By the above discussion, we
have a maximal $BPICTL$-consistent set $\Gamma $ such that $\Delta \subseteq
\Gamma $.%
%TCIMACRO{\TeXButton{End Proof}{\endproof}}%
%BeginExpansion
\endproof%
%EndExpansion

\textbf{Lemma 3. }$D_{a}$ satisfy the conditions of Definition of $BPICTL$
model $M.$

$Proof.$ By the construction of $BPICTL$ canonical model $M,$ for any $Q\in
D_{a},$ there is a $\varphi ,$ $Q=[[\varphi ]]_{M}.$ Therefore $D_{a}$
satisfy the conditions of Definition of $BPICTL$ model $M.$%
%TCIMACRO{\TeXButton{End Proof}{\endproof}}%
%BeginExpansion
\endproof%
%EndExpansion

In the following, we prove that $R_{a}^{B},R_{a}^{P},R_{a}^{I},R_{a}^{X}$
satisfy the conditions of Definition of $BPICTL$ model $M.$

\textbf{Lemma 4 }(B3). $(\forall \Gamma _{1}.\forall \Gamma _{2}.\forall
\Gamma _{3}.((\Gamma _{1},\Gamma _{2})\in R_{a}^{B}\wedge (\Gamma
_{2},\Gamma _{3})\in R_{a}^{B}\rightarrow (\Gamma _{1},\Gamma _{3})\in
R_{a}^{B})).$

$Proof.$ By $B_{a}\varphi \rightarrow B_{a}B_{a}\varphi .$%
%TCIMACRO{\TeXButton{End Proof}{\endproof}}%
%BeginExpansion
\endproof%
%EndExpansion

\textbf{Lemma 5} (B4). $(\forall \Gamma _{1}.\forall \Gamma _{2}.\forall
\Gamma _{3}.((\Gamma _{1},\Gamma _{2})\in R_{a}^{B}\wedge (\Gamma
_{1},\Gamma _{3})\in R_{a}^{B}\rightarrow (\Gamma _{2},\Gamma _{3})\in
R_{a}^{B})).$

$Proof.$ By $\lnot B_{a}\varphi \rightarrow B_{a}\lnot B_{a}\varphi .$%
%TCIMACRO{\TeXButton{End Proof}{\endproof}}%
%BeginExpansion
\endproof%
%EndExpansion

\textbf{Lemma 6} (B5). $(\forall \Gamma _{1}.\exists \Gamma _{2}.(\Gamma
_{1},\Gamma _{2})\in R_{a}^{B}).$

$Proof.$ By $B_{a}\varphi \rightarrow \lnot B_{a}\lnot \varphi .$%
%TCIMACRO{\TeXButton{End Proof}{\endproof}}%
%BeginExpansion
\endproof%
%EndExpansion

\textbf{Lemma 7} (P1). $\forall \Gamma .\forall Q_{1},Q_{2}.(Q_{1}\in
R_{a}^{P}(\Gamma )\wedge Q_{2}\in R_{a}^{P}(\Gamma )\rightarrow Q_{1}\cap
Q_{2}\in R_{a}^{P}(\Gamma )).$

$Proof.$ For any $\varphi _{1},\varphi _{2},\Gamma ,$ suppose $[[\varphi
_{1}]]_{M}\in R_{a}^{P}(\Gamma )\wedge \lbrack \lbrack \varphi _{2}]]_{M}\in
R_{a}^{P}(\Gamma ).$ Since $(P_{a}\varphi _{1}\wedge P_{a}\varphi
_{2})\rightarrow P_{a}(\varphi _{1}\wedge \varphi _{2}),$ we have $[[\varphi
_{1}]]_{M}\cap \lbrack \lbrack \varphi _{2}]]_{M}\in R_{a}^{P}(\Gamma ).$%
%TCIMACRO{\TeXButton{End Proof}{\endproof}}%
%BeginExpansion
\endproof%
%EndExpansion

\textbf{Lemma 8} (P2). $\forall \Gamma .\forall Q_{1},Q_{2}.(Q_{1}\in
R_{a}^{P}(\Gamma )\wedge \overline{Q_{1}}\cup Q_{2}\in R_{a}^{P}(\Gamma
)\rightarrow Q_{2}\in R_{a}^{P}(\Gamma )).$

$Proof.$ For any $\varphi _{1},\varphi _{2},\Gamma ,$ suppose $[[\varphi
_{1}]]_{M}\in R_{a}^{P}(\Gamma )\wedge \overline{[[\varphi _{1}]]_{M}}\cup
\lbrack \lbrack \varphi _{2}]]_{M}\in R_{a}^{P}(\Gamma ).$ Since $%
(P_{a}\varphi _{1}\wedge P_{a}\varphi _{1}\rightarrow \varphi
_{2})\rightarrow P_{a}(\varphi _{2}),$ we have $[[\varphi _{2}]]_{M}\in
R_{a}^{P}(\Gamma ).$%
%TCIMACRO{\TeXButton{End Proof}{\endproof}}%
%BeginExpansion
\endproof%
%EndExpansion

\textbf{Lemma 9} (P3). $\forall \Gamma _{1}.\forall Q.(\{\Gamma _{2}$ $|$ $%
Q\in R_{a}^{P}(\Gamma _{2})\}\in R_{a}^{P}(\Gamma _{1}))\rightarrow Q\in
R_{a}^{P}(\Gamma _{1}).$

$Proof.$ For any $\Gamma _{1}$, $[[\varphi ]]_{M},$ suppose $\{\Gamma _{2}$ $%
|$ $[[\varphi ]]_{M}\in R_{a}^{P}(\Gamma _{2})\}\in R_{a}^{P}(\Gamma _{1}).$
Therefore $[[P_{a}\varphi ]]_{M}\in R_{a}^{P}(\Gamma _{1}),$ $%
P_{a}P_{a}\varphi \in \Gamma _{1}.$ Since\ $(P_{a}P_{a}\varphi \rightarrow
P_{a}\varphi ),$ $P_{a}\varphi \in \Gamma _{1}.$ So $[[\varphi ]]_{M}\in
R_{a}^{P}(\Gamma _{1}).$%
%TCIMACRO{\TeXButton{End Proof}{\endproof}}%
%BeginExpansion
\endproof%
%EndExpansion

\textbf{Lemma 10} (P4). $\forall \Gamma _{1}.\forall Q_{1}.Q_{1}\in
R_{a}^{P}(\Gamma _{1})\rightarrow \forall Q_{2}.\exists \Gamma
_{2}.((Q_{2}\in R_{a}^{P}(\Gamma _{1})\rightarrow (\Gamma _{2}\in
Q_{2}\wedge Q_{1}\in R_{a}^{P}(\Gamma _{2}))).$

$Proof.$ \ Suppose\ $[[\varphi ]]_{M}\in R_{a}^{P}(\Gamma _{1}),$ $\Gamma
_{1}\in \lbrack \lbrack P_{a}\varphi ]]_{M}.$ Since $P_{a}\lnot P_{a}\varphi
\rightarrow \lnot P_{a}\varphi ,$ $P_{a}\varphi \rightarrow \lnot P_{a}\lnot
P_{a}\varphi .$ We have $\Gamma _{1}\in \lbrack \lbrack \lnot P_{a}\lnot
P_{a}\varphi ]]_{M}.$ Therefore for any $\psi ,$ $\exists \Gamma
_{2}.([[\psi ]]_{M}\in R_{a}^{P}(\Gamma _{1})\rightarrow (\Gamma _{2}\in
\lbrack \lbrack \psi ]]_{M}\wedge $ $\Gamma _{2}\in \lbrack \lbrack
P_{a}\varphi ]]_{M}).$%
%TCIMACRO{\TeXButton{End Proof}{\endproof}}%
%BeginExpansion
\endproof%
%EndExpansion

\textbf{Lemma 11} (BP1). $\forall \Gamma _{1}.\forall Q_{1},Q_{2}\in
D_{a}.(Q_{1}\in R_{a}^{P}(\Gamma _{1})\wedge \{\Gamma _{2}$ $|$ $(\Gamma
_{1},\Gamma _{2})\in R_{a}^{B}\}\subseteq (Q_{1}\cap Q_{2})\cup (\overline{%
Q_{1}}\cap \overline{Q_{2}}))\rightarrow Q_{2}\in R_{a}^{P}(\Gamma _{1}).$

$Proof.$ Suppose $Q_{1}\in R_{a}^{P}(\Gamma _{1})\wedge \{\Gamma _{2}$ $|$ $%
(\Gamma _{1},\Gamma _{2})\in R_{a}^{B}\}\subseteq (Q_{1}\cap Q_{2})\cup (%
\overline{Q_{1}}\cap \overline{Q_{2}}),\ $Therefore for any $\varphi _{1},$ $%
\varphi _{2},$ if $[[\varphi _{1}]]_{M}\in R_{a}^{P}(\Gamma _{1})\wedge
\{\Gamma _{2}$ $|$ $(\Gamma _{1},\Gamma _{2})\in R_{a}^{B}\}\subseteq
([[\varphi _{1}]]_{M}\cap \lbrack \lbrack \varphi _{2}]]_{M})\cup (\overline{%
[[\varphi _{1}]]_{M}}\cap \overline{[[\varphi _{2}]]_{M}}),$ by $%
(B_{a}(\varphi _{1}\leftrightarrow \varphi _{2})\wedge P_{a}\varphi
_{1})\rightarrow P_{a}\varphi _{2}),$ we have $B_{a}(\varphi
_{1}\leftrightarrow \varphi _{2})\wedge P_{a}\varphi _{1}\in \Gamma _{1},$
hence $P_{a}\varphi _{2}\in \Gamma _{1}.$ Therefore $Q_{2}=[[\varphi
_{2}]]_{M}\in R_{a}^{P}(\Gamma _{1}).$%
%TCIMACRO{\TeXButton{End Proof}{\endproof}}%
%BeginExpansion
\endproof%
%EndExpansion

\textbf{Lemma 12} (BP2). $\forall \Gamma _{1}.\forall Q.Q\in
R_{a}^{P}(\Gamma _{1})\rightarrow (\forall \Gamma _{2}.(\Gamma _{1},\Gamma
_{2})\in R_{a}^{B}\rightarrow Q\in R_{a}^{P}(\Gamma _{2}))$

$Proof.$ For any $\Gamma _{1},$ for any $\varphi ,$ if $Q=[[\varphi
]]_{M}\in R_{a}^{P}(\Gamma _{1}),$ then $P_{a}\varphi \in \Gamma _{1}.$ By $%
P_{a}\varphi \rightarrow B_{a}P_{a}\varphi $, if $P_{a}\varphi \in \Gamma
_{1},$ then $B_{a}P_{a}\varphi \in \Gamma _{1}.$ For any $\Gamma _{2},$ if $%
(\Gamma _{1},\Gamma _{2})\in R_{a}^{B},$ then $P_{a}\varphi \in \Gamma _{2},$%
and $Q=[[\varphi ]]_{M}\in R_{a}^{P}(\Gamma _{2}).$%
%TCIMACRO{\TeXButton{End Proof}{\endproof}}%
%BeginExpansion
\endproof%
%EndExpansion

\textbf{Lemma 13} (BP3). $\forall \Gamma _{1}.\forall Q\in D_{a}.(\exists
\Gamma _{2}.(\Gamma _{1},\Gamma _{2})\in R_{a}^{B}\wedge Q\in
R_{a}^{P}(\Gamma _{2}))\rightarrow Q\in R_{a}^{P}(\Gamma _{1}).$

$Proof.$ Suppose $\exists \Gamma _{2}.(\Gamma _{1},\Gamma _{2})\in
R_{a}^{B}\wedge \lbrack \lbrack \varphi ]]_{M}\in R_{a}^{P}(\Gamma _{2}),$
Therefore $P_{a}\varphi \in \Gamma _{2},$ $B_{a}\lnot P_{a}\varphi \notin
\Gamma _{1},$ $\lnot B_{a}\lnot P_{a}\varphi \in \Gamma _{1}.$ Since $\lnot
P_{a}\varphi \rightarrow B_{a}\lnot P_{a}\varphi $, we have $\lnot
B_{a}\lnot P_{a}\varphi \rightarrow P_{a}\varphi .$ Therefore $P_{a}\varphi
\in \Gamma _{1}.$ Therefore $[[\varphi ]]_{M}\in R_{a}^{P}(\Gamma _{1}).$%
%TCIMACRO{\TeXButton{End Proof}{\endproof}}%
%BeginExpansion
\endproof%
%EndExpansion

\textbf{Lemma 14} (BP4). $\forall \Gamma _{1}.\forall Q.(\forall \Gamma
_{2}.(\Gamma _{1},\Gamma _{2})\in R_{a}^{B}\rightarrow Q\in R_{a}^{P}(\Gamma
_{2}))\rightarrow Q\in R_{a}^{P}(\Gamma _{1})$

$Proof.$ \ Suppose for any $\varphi ,$ $\forall \Gamma _{1}.(\forall \Gamma
_{2}.(\Gamma _{1},\Gamma _{2})\in R_{a}^{B}\rightarrow \lbrack \lbrack
\varphi ]]_{M}\in R_{a}^{P}(\Gamma _{2})),$ so $\Gamma _{2}\in \lbrack
\lbrack P_{a}\varphi ]]_{M},\Gamma _{1}\in \lbrack \lbrack B_{a}P_{a}\varphi
]]_{M}.$ Since $B_{a}P_{a}\varphi \rightarrow P_{a}\varphi ,$ $\Gamma
_{1}\in \lbrack \lbrack P_{a}\varphi ]]_{M}.$ Therefore $[[\varphi ]]_{M}\in
R_{a}^{P}(\Gamma _{1}).$%
%TCIMACRO{\TeXButton{End Proof}{\endproof}}%
%BeginExpansion
\endproof%
%EndExpansion

\textbf{Lemma 15} (BP5). $\forall \Gamma _{1}.\forall Q\in R_{a}^{P}(\Gamma
_{1})\rightarrow \exists \Gamma _{2}.((\Gamma _{1},\Gamma _{2})\in
R_{a}^{B}\wedge Q\in R_{a}^{P}(\Gamma _{2})).$

$Proof.$ \ Suppose\ $[[\varphi ]]_{M}\in R_{a}^{P}(\Gamma _{1}),$ then $%
[[P_{a}\varphi ]]_{M}\in \Gamma _{1}.$ Since $B_{a}\lnot P_{a}\varphi
\rightarrow \lnot P_{a}\varphi ,$ $P_{a}\varphi \rightarrow \lnot B_{a}\lnot
P_{a}\varphi .$ We have $[[\lnot B_{a}\lnot P_{a}\varphi ]]_{M}\in \Gamma
_{1}.$ There is $\Gamma _{2},$ $(\Gamma _{1},\Gamma _{2})\in R_{a}^{B},$ $%
[[P_{a}\varphi ]]_{M}\in \Gamma _{2}.$ Therefore $[[\varphi ]]_{M}\in
R_{a}^{P}(\Gamma _{2}).$%
%TCIMACRO{\TeXButton{End Proof}{\endproof}}%
%BeginExpansion
\endproof%
%EndExpansion

\textbf{Lemma 16} (BI1). $\forall \Gamma _{1}.\forall Q_{1},Q_{2}\in
D_{a}.(Q_{1}\in R_{a}^{I}(\Gamma _{1})\wedge \{\Gamma _{2}$ $|$ $(\Gamma
_{1},\Gamma _{2})\in R_{a}^{B}\}\subseteq (Q_{1}\cap Q_{2})\cup (\overline{%
Q_{1}}\cap \overline{Q_{2}}))\rightarrow Q_{2}\in R_{a}^{I}(\Gamma _{1}).$

$Proof.$ Suppose $Q_{1}\in R_{a}^{I}(\Gamma _{1})\wedge \{\Gamma _{2}$ $|$ $%
(\Gamma _{1},\Gamma _{2})\in R_{a}^{B}\}\subseteq (Q_{1}\cap Q_{2})\cup (%
\overline{Q_{1}}\cap \overline{Q_{2}}),\ $Therefore for any $\varphi _{1},$ $%
\varphi _{2},$ if $[[\varphi _{1}]]_{M}\in R_{a}^{I}(\Gamma _{1})\wedge
\{\Gamma _{2}$ $|$ $(\Gamma _{1},\Gamma _{2})\in R_{a}^{B}\}\subseteq
([[\varphi _{1}]]_{M}\cap \lbrack \lbrack \varphi _{2}]]_{M})\cup (\overline{%
[[\varphi _{1}]]_{M}}\cap \overline{[[\varphi _{2}]]_{M}}),$ by $%
(B_{a}(\varphi _{1}\leftrightarrow \varphi _{2})\wedge I_{a}\varphi
_{1})\rightarrow I_{a}\varphi _{2}),$ we have $B_{a}(\varphi
_{1}\leftrightarrow \varphi _{2})\wedge I_{a}\varphi _{1}\in \Gamma _{1},$
hence $I_{a}\varphi _{2}\in \Gamma _{1}.$ Therefore $Q_{2}=[[\varphi
_{2}]]_{M}\in R_{a}^{I}(\Gamma _{1}).$%
%TCIMACRO{\TeXButton{End Proof}{\endproof}}%
%BeginExpansion
\endproof%
%EndExpansion

\textbf{Lemma 17} (BI2). $\forall \Gamma _{1}.\forall Q.Q\in
R_{a}^{I}(\Gamma _{1})\rightarrow (\forall \Gamma _{2}.(\Gamma _{1},\Gamma
_{2})\in R_{a}^{B}\rightarrow Q\in R_{a}^{I}(\Gamma _{2}))$

$Proof.$ For any $\Gamma _{1},$ for any $\varphi ,$ if $Q=[[\varphi
]]_{M}\in R_{a}^{I}(\Gamma _{1}),$ then $I_{a}\varphi \in \Gamma _{1}.$ By $%
I_{a}\varphi \rightarrow B_{a}I_{a}\varphi $, if $I_{a}\varphi \in \Gamma
_{1},$ then $B_{a}I_{a}\varphi \in \Gamma _{1}.$ Suppose $(\Gamma
_{1},\Gamma _{2})\in R_{a}^{B},$ then $I_{a}\varphi \in \Gamma _{2},$and $%
Q=[[\varphi ]]_{M}\in R_{a}^{I}(\Gamma _{2}).$%
%TCIMACRO{\TeXButton{End Proof}{\endproof}}%
%BeginExpansion
\endproof%
%EndExpansion

\textbf{Lemma 18} (BI3). $\forall \Gamma _{1}.\forall Q\in D_{a}.(\exists
\Gamma _{2}.(\Gamma _{1},\Gamma _{2})\in R_{a}^{B}\wedge Q\in
R_{a}^{I}(\Gamma _{2}))\rightarrow Q\in R_{a}^{I}(\Gamma _{1}).$

$Proof.$ Suppose $\exists \Gamma _{2}.(\Gamma _{1},\Gamma _{2})\in
R_{a}^{B}\wedge \lbrack \lbrack \varphi ]]_{M}\in R_{a}^{I}(\Gamma _{2}),$
Therefore $\lnot B_{a}\lnot I_{a}\varphi \in \Gamma _{1}.$ Since $\lnot
I_{a}\varphi \rightarrow B_{a}\lnot I_{a}\varphi $, we have $\lnot
B_{a}\lnot I_{a}\varphi \rightarrow I_{a}\varphi .$ Therefore $I_{a}\varphi
\in \Gamma _{1}.$ Therefore $[[\varphi ]]_{M}\in R_{a}^{I}(\Gamma _{1}).$%
%TCIMACRO{\TeXButton{End Proof}{\endproof}}%
%BeginExpansion
\endproof%
%EndExpansion

\textbf{Lemma 19} (BI4). $\forall \Gamma _{1}.\forall Q_{1}.(\forall \Gamma
_{2}.(\Gamma _{1},\Gamma _{2})\in R_{a}^{B}\rightarrow Q_{1}\in
R_{a}^{I}(\Gamma _{2}))\rightarrow Q_{1}\in R_{a}^{I}(\Gamma _{1}).$

$Proof.$ \ Suppose for any $\varphi ,$ $\forall \Gamma _{1}.(\forall \Gamma
_{2}.(\Gamma _{1},\Gamma _{2})\in R_{a}^{B}\rightarrow \lbrack \lbrack
\varphi ]]_{M}\in R_{a}^{I}(\Gamma _{2})),$ so $\Gamma _{2}\in \lbrack
\lbrack I_{a}\varphi ]]_{M},\Gamma _{1}\in \lbrack \lbrack B_{a}I_{a}\varphi
]]_{M}.$ Since $B_{a}I_{a}\varphi \rightarrow I_{a}\varphi ,$ $\Gamma
_{1}\in \lbrack \lbrack I_{a}\varphi ]]_{M}.$ Therefore $[[\varphi ]]_{M}\in
R_{a}^{I}(\Gamma _{1}).$%
%TCIMACRO{\TeXButton{End Proof}{\endproof}}%
%BeginExpansion
\endproof%
%EndExpansion

\textbf{Lemma 20 }(BI5).\textbf{\ }$\forall \Gamma _{1}.\forall Q\in
R_{a}^{I}(\Gamma _{1})\rightarrow \exists \Gamma _{2}.((\Gamma _{1},\Gamma
_{2})\in R_{a}^{B}\wedge Q\in R_{a}^{I}(\Gamma _{2})).$

$Proof.$ \ Suppose\ $[[\varphi ]]_{M}\in R_{a}^{I}(\Gamma _{1}),$ then $%
[[I_{a}\varphi ]]_{M}\in \Gamma _{1}.$ Since $B_{a}\lnot I_{a}\varphi
\rightarrow \lnot I_{a}\varphi ,$ $I_{a}\varphi \rightarrow \lnot B_{a}\lnot
I_{a}\varphi .$ We have $[[\lnot B_{a}\lnot I_{a}\varphi ]]_{M}\in \Gamma
_{1}.$ There is $\Gamma _{2},$ $(\Gamma _{1},\Gamma _{2})\in R_{a}^{B},$ $%
[[I_{a}\varphi ]]_{M}\in \Gamma _{2}.$ Therefore $[[\varphi ]]_{M}\in
R_{a}^{I}(\Gamma _{2}).$%
%TCIMACRO{\TeXButton{End Proof}{\endproof}}%
%BeginExpansion
\endproof%
%EndExpansion

\textbf{Lemma 21} (BPIEF1a). $\forall \Gamma .\forall Q.(Q\in
R_{a}^{I}(\Gamma )\rightarrow Q\in R_{a}^{P}(\Gamma )).$

$Proof.$ For any $\Gamma ,$ for any $\varphi ,$ suppose $[[\varphi ]]_{M}\in
R_{a}^{I}(\Gamma ).$ For any $\varphi ,$ since $I_{a}\varphi \rightarrow
P_{a}\varphi ,$ if $I_{a}\varphi \in \Gamma ,$ then $P_{a}\varphi \in \Gamma
.$ Hence $[[\varphi ]]_{M}\in R_{a}^{P}(\Gamma ).$ Therefore $\forall \Gamma
.\forall Q.(Q\in R_{a}^{I}(\Gamma )\rightarrow Q\in R_{a}^{P}(\Gamma )).$%
%TCIMACRO{\TeXButton{End Proof}{\endproof}}%
%BeginExpansion
\endproof%
%EndExpansion

\textbf{Lemma 22} (BPIEF1b). $\forall \Gamma _{1}.(\cup \{Q$ \TEXTsymbol{%
\vert} $Q\in R_{a}^{I}(\Gamma _{1}))\}\cap \{\Gamma _{2}$ \TEXTsymbol{\vert} 
$(\Gamma _{1},\Gamma _{2})\in R_{a}^{B}\}=\varnothing .$

$Proof.$ For any $\Gamma _{1},$ suppose $\Gamma _{2}\in \cup \{Q$ 
\TEXTsymbol{\vert} $Q\in R_{a}^{I}(\Gamma _{1}))\},$\ then $\exists \varphi
.I_{a}\varphi \in \Gamma _{1},$ and $\varphi \in \Gamma _{2}.$\ Since $%
I_{a}\varphi \rightarrow B_{a}\lnot \varphi $, we have $B_{a}\lnot \varphi
\in \Gamma _{1},$and $\varphi \in \Gamma _{2}.$ Therefore $\Gamma _{2}\notin
\{y$ \TEXTsymbol{\vert} $(\Gamma _{1},y)\in R_{a}^{B}\}.$ It holds that $%
\forall \Gamma _{1}.(\cup \{Q$ \TEXTsymbol{\vert} $Q\in R_{a}^{I}(\Gamma
_{1}))\}\cap \{\Gamma _{2}$ \TEXTsymbol{\vert} $(\Gamma _{1},\Gamma _{2})\in
R_{a}^{B}\}=\varnothing .$%
%TCIMACRO{\TeXButton{End Proof}{\endproof}}%
%BeginExpansion
\endproof%
%EndExpansion

\textbf{Lemma 23} (BPIEF1c). $\forall \Gamma _{1}.\forall \Gamma
_{2}.\forall Q\in D_{a}.((\Gamma _{1},\Gamma _{2})\in R_{a}^{B}\wedge Q\in
R_{a}^{I}(\Gamma _{1}))\rightarrow \exists \Gamma _{3}.((\Gamma _{2},\Gamma
_{3})\in R_{a}^{EF}\wedge \Gamma _{3}\in Q),$ where $R_{a}^{EF}$ is the
reflexive and transitive closure of $R_{a}^{X}.$

$Proof.$ For any $\Gamma _{1},$ for any $\varphi ,$ suppose $[[\varphi
]]_{M}\in R_{a}^{I}(\Gamma _{1}).$ By $I_{a}\varphi \rightarrow
B_{a}EF\varphi ,$ we have $\Gamma _{1}\in \lbrack \lbrack B_{a}EF\varphi
]]_{M}.$ Therefore $\forall \Gamma _{2}.(\Gamma _{1},\Gamma _{2})\in
R_{a}^{B}\rightarrow \exists \Gamma _{3}.((\Gamma _{2},\Gamma _{3})\in
R_{a}^{EF}\wedge \Gamma _{3}\in \lbrack \lbrack \varphi ]]_{M}).$ So $%
\forall \Gamma _{1}.\forall \Gamma _{2}.([[\varphi ]]_{M}\in
R_{a}^{I}(\Gamma _{1})\rightarrow ((\Gamma _{1},\Gamma _{2})\in
R_{a}^{B}\rightarrow \exists \Gamma _{3}.((\Gamma _{2},\Gamma _{3})\in
R_{a}^{EF}\wedge \Gamma _{3}\in \lbrack \lbrack \varphi ]]_{M}))),$ we have $%
\forall \Gamma _{1}.\forall \Gamma _{2}.((\Gamma _{1},\Gamma _{2})\in
R_{a}^{B}\wedge \lbrack \lbrack \varphi ]]_{M}\in R_{a}^{I}(\Gamma
_{1}))\rightarrow \exists \Gamma _{3}.((\Gamma _{2},\Gamma _{3})\in
R_{a}^{EF}\wedge \Gamma _{3}\in \lbrack \lbrack \varphi ]]_{M}).$ 
%TCIMACRO{\TeXButton{End Proof}{\endproof}}%
%BeginExpansion
\endproof%
%EndExpansion

\textbf{Lemma 24} (BX1). $\forall \Gamma _{1}.\{\Gamma _{2}$ $|$ $(\Gamma
_{1},\Gamma _{2})\in R_{a}^{B}\circ R_{a}^{X}\circ R_{a}^{B}\}\subseteq
\{\Gamma _{2}$ $|$ $(\Gamma _{1},\Gamma _{2})\in R_{a}^{B}\circ R_{a}^{X}\}.$

$Proof.$ For any $\Gamma _{1},$ suppose $(\Gamma _{1},\Gamma _{2})\in
R_{a}^{B}\circ R_{a}^{X}\circ R_{a}^{B}.$ Therefore there is $\Gamma _{3}$
such that $(\Gamma _{1},\Gamma _{3})\in R_{a}^{B}\circ R_{a}^{X},$ $(\Gamma
_{3},\Gamma _{2})\in R_{a}^{B}.$ Then for any $\varphi ,$ $\varphi \in
\Gamma _{2}$ implies $\lnot B_{a}\lnot \varphi \in \Gamma _{3},$ and $\lnot
B_{a}\lnot \varphi \in \Gamma _{3}$ implies $\lnot B_{a}AXB_{a}\lnot \varphi
\in \Gamma _{1}.$ By $B_{a}AX\varphi \rightarrow B_{a}AXB_{a}\varphi $, we
have $\lnot B_{a}AXB_{a}\lnot \varphi \rightarrow \lnot B_{a}AX\lnot \varphi
.$ Therefore for any $\varphi ,$ $\varphi \in \Gamma _{2}$ implies $\lnot
B_{a}AX\lnot \varphi \in \Gamma _{1}.$ So $B_{a}AX\lnot \varphi \in \Gamma
_{1}$ implies $\lnot \varphi \in \Gamma _{2},$ we have $(\Gamma _{1},\Gamma
_{2})\in R_{a}^{B}\circ R_{a}^{X}.$%
%TCIMACRO{\TeXButton{End Proof}{\endproof}}%
%BeginExpansion
\endproof%
%EndExpansion

\textbf{Lemma 25} (BX2). $\forall \Gamma _{1}.\forall Q\in D_{a}.(\forall
\Gamma _{2}.\exists \Gamma _{3}.(\Gamma _{1},\Gamma _{2})\in
R_{a}^{B}\rightarrow ((\Gamma _{2},\Gamma _{3})\in R_{a}^{X}\wedge \Gamma
_{3}\in Q))\rightarrow (\forall \Gamma _{4}.\exists \Gamma _{5}.\forall
\Gamma _{6}.(\Gamma _{1},\Gamma _{4})\in R_{a}^{B}\rightarrow ((\Gamma
_{4},\Gamma _{5})\in R_{a}^{X})\wedge ((\Gamma _{5},\Gamma _{6})\in
R_{a}^{B}\rightarrow \Gamma _{6}\in Q)).$

$Proof.$ For any $\Gamma _{1},$ any $\varphi ,$ suppose $\forall \Gamma
_{2}.\exists \Gamma _{3}.(\Gamma _{1},\Gamma _{2})\in R_{a}^{B}$ $%
\rightarrow ((\Gamma _{2},\Gamma _{3})\in R_{a}^{X}\wedge \Gamma _{3}\in
\lbrack \lbrack \varphi ]]_{M}),$ we have $B_{a}EX\varphi \in \Gamma _{1}.$%
By $B_{a}EX\varphi \rightarrow B_{a}EXB_{a}\varphi $, we have $%
B_{a}EXB_{a}\varphi \in \Gamma _{1}.$Therefore $\forall \Gamma _{4}.\exists
\Gamma _{5}.\forall \Gamma _{6}.(\Gamma _{1},\Gamma _{4})\in
R_{a}^{B}\rightarrow ((\Gamma _{4},\Gamma _{5})\in R_{a}^{X})\wedge ((\Gamma
_{5},\Gamma _{6})\in R_{a}^{B}\rightarrow \Gamma _{6}\in \lbrack \lbrack
\varphi ]]_{M})).$%
%TCIMACRO{\TeXButton{End Proof}{\endproof}}%
%BeginExpansion
\endproof%
%EndExpansion

\textbf{Lemma 26.} The canonical model $M$ is a $BPICTL$ model.

$Proof.$ It follows from Lemma 1 to Lemma 25.%
%TCIMACRO{%
%\TeXButton{End
%Proof}{\endproof}}%
%BeginExpansion
\endproof%
%EndExpansion

The above lemmas state that the model satisfies all conditions in Definition
2, then as a consequence, the model $M$ is a $BPICTL$ model. In order to get
the completeness, we further prove the following lemma, which states that $M$
is \textquotedblright canonical\textquotedblright .

\textbf{Lemma 27.} In the model $M$, for any $\Gamma $ and any $\varphi $, $%
(M,\Gamma )\models \varphi \Leftrightarrow \varphi \in \Gamma $.

$Proof.$ We argue by the cases on the structure of $\varphi $, here we only
give the proof in the cases of (1) $\varphi \equiv B_{a}\psi $ and (2) $%
\varphi \equiv I_{a}\psi $.

(1) It suffices to prove that: $(M,\Gamma )\models B_{a}\psi \Leftrightarrow
B_{a}\psi \in \Gamma $.

If $B_{a}\psi \in \Gamma $, by the definition of $M$, for every $(\Gamma
,\Gamma ^{\prime })\in R_{a}^{B}$, $\psi \in \Gamma ^{\prime },$ therefore $%
(M,\Gamma )\models B_{a}\psi $.

If $B_{a}\psi \notin \Gamma $, by the definition of $M$, we have $\lnot
B_{a}\psi \in \Gamma $. There is $\Gamma ^{\prime }=\{\lnot \psi \}\cup
\Gamma /B_{a}$ such that $\Gamma ^{\prime }$ is consistent and $(\Gamma
,\Gamma ^{\prime })\in R_{a}^{B}$, therefore $(M,\Gamma )\not\models
B_{a}\psi $.

(2) It suffices to prove that: $(M,\Gamma )\models I_{a}\psi \Leftrightarrow
I_{a}\psi \in \Gamma $.

If $I_{a}\psi \in \Gamma $, by the definition of $M$, $\{\Gamma ^{\prime }$ $%
|$ $\psi \in \Gamma ^{\prime }\}\in R_{a}^{I}(\Gamma ),$ therefore $%
(M,\Gamma )\models I_{a}\psi $.

If $I_{a}\psi \notin \Gamma $, by the definition of $M$, we have $\{\Gamma
^{\prime }$ $|$ $\psi \in \Gamma ^{\prime }\}\notin R_{a}^{I}(\Gamma ),$
therefore $(M,\Gamma )\not\models I_{a}\psi $.%
%TCIMACRO{\TeXButton{End Proof}{\endproof}}%
%BeginExpansion
\endproof%
%EndExpansion

Now it is ready to get the completeness of $BPICTL$:

\textbf{Proposition 2} (Completeness of $BPICTL$). The inference system of $%
BPICTL$ is complete, i.e., If $\Gamma \models _{BPICTL}\varphi $, then $%
\Gamma \vdash _{BPICTL}\varphi $.

$Proof.$ Suppose not, then there is a $BPICTL$ - consistent formulae set $%
\Phi =\Gamma \cup \{\lnot \varphi \}$, and there is no model $M$ such that $%
\Phi $ is satisfied in $M$. For there is a $BPICTL$ - maximal consistent
formula set $\Sigma $ such that $\Phi \subseteq \Sigma $, by Lemma 27, $\Phi 
$ is satisfied in possible world $\Sigma $ of $M$. It is a contradiction.%
%TCIMACRO{\TeXButton{End Proof}{\endproof}}%
%BeginExpansion
\endproof%
%EndExpansion

Proposition 1 and Proposition 2 show that the axioms and inference rules of $%
BPICTL$ give us a sound and complete axiomatization for belief, preference,
intention and temporal property.

In the following, we give some corollaries of the inference system of $%
BPICTL.$

\textbf{Corollary 1.}

(1) $\vdash (\varphi _{1}\leftrightarrow \varphi _{2})\Rightarrow \vdash
P_{a}\varphi _{1}\leftrightarrow P_{a}\varphi _{2}.$

(2) $\vdash (\varphi _{1}\leftrightarrow \varphi _{2})\Rightarrow \vdash
I_{a}\varphi _{1}\leftrightarrow I_{a}\varphi _{2}.$

$Proof.$ (1) Suppose $\vdash (\varphi _{1}\leftrightarrow \varphi _{2}),$ we
have $\vdash B_{a}(\varphi _{1}\leftrightarrow \varphi _{2}).$ Since $%
(B_{a}(\varphi _{1}\leftrightarrow \varphi _{2})\wedge P_{a}\varphi
_{1})\rightarrow P_{a}\varphi _{2},$ we have $\vdash P_{a}\varphi
_{1}\rightarrow P_{a}\varphi _{2}.$ Similarly, we have $\vdash P_{a}\varphi
_{2}\rightarrow P_{a}\varphi _{1}.$ Therefore $\vdash P_{a}\varphi
_{1}\leftrightarrow P_{a}\varphi _{2}.$

(2) Suppose $\vdash (\varphi _{1}\leftrightarrow \varphi _{2}),$ we have $%
\vdash B_{a}(\varphi _{1}\leftrightarrow \varphi _{2}).$ Since $%
(B_{a}(\varphi _{1}\leftrightarrow \varphi _{2})\wedge I_{a}\varphi
_{1})\rightarrow I_{a}\varphi _{2},$ we have $\vdash I_{a}\varphi
_{1}\rightarrow I_{a}\varphi _{2}.$ Similarly, we have $\vdash I_{a}\varphi
_{2}\rightarrow I_{a}\varphi _{1}.$ Therefore $\vdash I_{a}\varphi
_{1}\leftrightarrow I_{a}\varphi _{2}.$%
%TCIMACRO{\TeXButton{End Proof}{\endproof}}%
%BeginExpansion
\endproof%
%EndExpansion

\textbf{Corollary 2.} $(P_{a}\varphi _{1}\wedge P_{a}(\varphi
_{1}\rightarrow P_{a}\varphi _{2}))\rightarrow P_{a}\varphi _{2}.$

$Proof.$ Since $(P_{a}\varphi _{1}\wedge P_{a}(\varphi _{1}\rightarrow
P_{a}\varphi _{2})),$ by $(P_{a}\varphi _{1}\wedge P_{a}(\varphi
_{1}\rightarrow \varphi _{2}))\rightarrow P_{a}(\varphi _{2}),$ we have $%
P_{a}P_{a}\varphi _{2}.$ Since $P_{a}P_{a}\varphi _{2}\rightarrow
P_{a}\varphi _{2},$ we have $P_{a}\varphi _{2}.$%
%TCIMACRO{\TeXButton{End Proof}{\endproof}}%
%BeginExpansion
\endproof%
%EndExpansion

\section{Finite Model Property of $BPICTL$}

We now turn our attention to the finite model property of $BPICTL$. It needs
to show that if a formula is $BPICTL$-consistent, then it is satisfiable in
a finite model. The idea is that rather than considering maximal consistent
formulae set when trying to construct a model satisfying a formula $\varphi $%
, we restrict our attention to sets of subformulae of $\varphi $.

\textbf{Definition 5.} A finite model $M$ of $BPICTL$ is a structure $M=(S,$ 
$\Pi ,$ $\pi ,$ $R_{a}^{B},$ $R_{a}^{P},$ $R_{a}^{I},$ $R_{a}^{X},D_{a})$,
where

(1) $S$ is a nonempty finite set, whose elements are called possible worlds
or states.

(2) $\Pi $ is a nonempty set of atomic formulas.

(3) $\pi $ is a map: $S\times \Pi \longrightarrow \{true,false\}$.

(4) $R_{a}^{B}:R_{a}^{B}\subseteq S\times S$ is an accessible relation on $S$%
, which is a belief relation.

(5) $R_{a}^{P}:S\longrightarrow \wp (\wp (S))$, where $\wp (S)$ is the power
set of $S$.

(6) $R_{a}^{I}:S\longrightarrow \wp (\wp (S)).$

(7) $R_{a}^{X}:R_{a}^{X}\subseteq S\times S$ is an accessible relation on $S$%
, which is a temporal relation.

(8) $D_{a}=\wp (S).$

The conditions of $R_{a}^{B},R_{a}^{P},R_{a}^{I},R_{a}^{X}$ is same as in
Definition 2, except that $D_{a}=\wp (S)$ here.

\textbf{Definition 6.} Suppose $\zeta $ is a consistent formula with respect
to $BPICTL$, $Sub^{\ast }(\zeta )$ is a set of formulae defined as follows:
let $\zeta \in L^{BPICTL}$, $Sub(\zeta )$ is the set of subformulae of $%
\zeta $, then $Sub^{\ast }(\zeta )=Sub(\zeta )\cup \{\lnot \psi |\psi \in
Sub(\zeta )\}$. It is clear that $Sub^{\ast }(\zeta )$ is finite.

\textbf{Definition 7.} The finite canonical model $M_{\zeta }$ with respect
to $BPICTL$ formula $\zeta $ is $(S_{\zeta },\Pi _{\zeta },\pi _{\zeta
},R_{a,\zeta }^{B},R_{a,\zeta }^{P},R_{a,\zeta }^{I},R_{a,\zeta
}^{X},D_{a,\zeta })$.

(1) $S_{\zeta }=\{\Gamma $ $|$ $\Gamma $ is a maximal consistent set with
respect to $Sub^{\ast }(\zeta ),$ i.e., $\Gamma \subseteq Sub^{\ast }(\zeta
),$ $\Gamma $ is consistent, and $\Psi \in \Gamma $ or $\lnot \Psi \in
\Gamma $ for every $\Psi \in Sub^{\ast }(\zeta ).\}$.

(2) $\Pi _{\zeta }$ is the set of atomic formulae.

(3) $\pi _{\zeta }$ is a truth assignment as follows: for any atomic formula 
$p$, $\pi _{\zeta }(\Gamma ,p)=true\Leftrightarrow p\in \Gamma $.

(4) $R_{a,\zeta }^{B}=\{(\Gamma _{1},\Gamma _{2})$ $|$ $\Gamma
_{1}/B_{a}\subseteq \Gamma _{2},$ $\Gamma _{1}\in S_{\zeta },$ $\Gamma
_{2}\in S_{\zeta }\},$ where $\Gamma _{1}/B_{a}=\{\varphi $ $|$ $%
B_{a}\varphi \in \Gamma _{1}\}.$

(5) $R_{a,\zeta }^{P}$ maps every element of $S$ to a subset of $\wp (S)$: $%
R_{a}^{P}(\Gamma )=\{U_{a}^{P}(\varphi )$ $|$ $\varphi $ is a formula of $%
Sub^{\ast }(\zeta )$ and $P_{a}\varphi \in \Gamma ,$ where $%
U_{a}^{P}(\varphi )=\{\Gamma ^{\prime }$ $|$ $\varphi \in \Gamma ^{\prime },$%
and $P_{a}\varphi \in \Gamma ,\Gamma \in S_{\zeta },\Gamma ^{\prime }\in
S_{\zeta }\}\}$.

(6) $R_{a,\zeta }^{I}$ maps every element of $S$ to a subset of $\wp (S)$: $%
R_{a}^{I}(\Gamma )=\{U_{a}^{I}(\varphi )$ $|$ $\varphi $ is a formula of $%
Sub^{\ast }(\zeta )$ and $I_{a}\varphi \in \Gamma ,$ where $%
U_{a}^{I}(\varphi )=\{\Gamma ^{\prime }$ $|$ $\varphi \in \Gamma ^{\prime },$%
and $I_{a}\varphi \in \Gamma ,\Gamma \in S_{\zeta },\Gamma ^{\prime }\in
S_{\zeta }\}\}$.

(7) $R_{a,\zeta }^{X}=\{(\Gamma _{1},\Gamma _{2})$ $|$ $\Gamma
_{1}/AX_{a}\subseteq \Gamma _{2},$ $\Gamma _{1}\in S_{\zeta },\Gamma _{2}\in
S_{\zeta }\},$ where $\Gamma _{1}/AX_{a}=\{\varphi $ $|$ $AX\varphi \in
\Gamma _{1}\}.$

(8) $D_{a,\zeta }=\{[[\varphi ]]_{M_{\zeta }}$ \TEXTsymbol{\vert} $\varphi $
is a $BPICTL$ formula$\}.$

Similar to the proof of completeness of $BPICTL$, we mainly need to show
that the above canonical model $M_{\zeta }$ is a $BPICTL$ model. The
following lemmas contribute to this purpose.

\textbf{Lemma 28.} $S_{\zeta }$ is a nonempty finite set.

$Proof.$ Since the rules and axioms of $BPICTL$ are consistent, $S_{\zeta }$
is nonempty. For $Sub^{\ast }(\zeta )$ is a finite set, by the definition of 
$S_{\zeta }$, the cardinality of $S_{\zeta }$ is no more than the
cardinality of $\wp (Sub^{\ast }(\zeta ))$.%
%TCIMACRO{\TeXButton{End Proof}{\endproof}}%
%BeginExpansion
\endproof%
%EndExpansion

\textbf{Lemma 29.} $D_{a,\zeta }=\{[[\varphi ]]_{M_{\zeta }}$ \TEXTsymbol{%
\vert} $\varphi $ is a $BPICTL$ formula$\}$ is the power set of $S_{\zeta }$.

$Proof.$ Firstly, since $Sub^{\ast }(\zeta )$ is finite, so if $\Gamma \in
S_{\zeta }$ then $\Gamma $ is finite. We can let $\varphi _{\Gamma }$ be the
conjunction of the formulae in $\Gamma $. Secondly, if $A\subseteq S_{\zeta
} $, then $A=[[(\vee _{\Gamma \in A}\varphi _{\Gamma })]]_{M_{\zeta }}$. By
the above argument, we have that $D_{a,\zeta }$ is the power set of $%
S_{\zeta }$.%
%TCIMACRO{\TeXButton{End Proof}{\endproof}}%
%BeginExpansion
\endproof%
%EndExpansion

\textbf{Lemma 30.} If $\varphi $ is consistent (here $\varphi $ is a Boolean
combination of formulae in $Sub^{\ast }(\zeta )$), then there exists $\Gamma 
$ such that $\varphi $ can be proved from $\Gamma $, here $\Gamma $ is a
maximal consistent set with respect to $Sub^{\ast }(\zeta )$.

$Proof.$ For $\varphi $ is a Boolean combination of formulae in $Sub^{\ast
}(\zeta )$, therefore by regarding the formulae in $Sub^{\ast }(\zeta )$ as
atomic formulae, $\varphi $ can be represented as disjunctive normal form.
Since $\varphi $ is consistent, so there is a consistent disjunctive term in
disjunctive normal form expression of $\varphi $, let such term be $\psi
_{1}\wedge $...$\wedge \psi _{n}$, then $\varphi $ can be derived from the
maximal consistent set $\Gamma $ which contains $\{\psi _{1},...,\psi _{n}\}$%
.%
%TCIMACRO{\TeXButton{End Proof}{\endproof}}%
%BeginExpansion
\endproof%
%EndExpansion

\textbf{Lemma 31.} The model $M_{\zeta }$ is a $BPICTL$-model.

$Proof.$ Similar to Lemma 26\textbf{.}%
%TCIMACRO{\TeXButton{End Proof}{\endproof}}%
%BeginExpansion
\endproof%
%EndExpansion

\textbf{Lemma 32.} The model $M_{\zeta }$ is a finite model.

$Proof.$ By the definition of $S_{\zeta }$, the cardinality of $S_{\zeta }$
is no more than the cardinality of $\wp (Sub^{\ast }(\zeta ))$, which means $%
|S_{\zeta }|\leq 2^{|Sub^{\ast }(\zeta )|}$. 
%TCIMACRO{\TeXButton{End Proof}{\endproof}}%
%BeginExpansion
\endproof%
%EndExpansion

Similar to the proof of completeness of $BPICTL$, the above lemmas show that 
$M_{\zeta }$ is a finite $BPICTL$-model and the following lemma states that $%
M_{\zeta }$ is canonical.

\textbf{Lemma 33.} For the finite canonical model $M_{\zeta }$, for any $%
\Gamma \in S_{\zeta }$ and any $\varphi \in Sub^{\ast }(\zeta )$, $(M_{\zeta
},\Gamma )\models \varphi \Leftrightarrow \varphi \in \Gamma $.

$Proof.$ Similar to Lemma 27\textbf{.}%
%TCIMACRO{\TeXButton{End Proof}{\endproof}}%
%BeginExpansion
\endproof%
%EndExpansion

From the above lemmas, we know that $M_{\zeta }$ is a finite $BPICTL$-model
that is canonical. Now it is not difficult to get the following proposition.

\textbf{Proposition 3} (Finite model property of $BPICTL$). If $\Gamma $ is
a finite set of formulae and $\wedge \Gamma $ is consistent, then there is a
finite $BPICTL$-model $M$ and a state $s$ such that $(M,s)\models \Gamma $.

$Proof.$ By Lemma 33, there exists a finite $BPICTL$-model $M_{\wedge \Gamma
}$ such that $\Gamma $ is satisfied in $M_{\wedge \Gamma }$.%
%TCIMACRO{\TeXButton{End Proof}{\endproof}}%
%BeginExpansion
\endproof%
%EndExpansion

Usually, in the case of modal logics, one can get decidability of the
provability problem from finite model property. At first, one can simply
construct every model with finite (for example, say $2^{|Sub^{\ast }(\varphi
)|}$) states. One then check if $\varphi $ is true at some state of one of
these models (note that the number of models that have $2^{|Sub^{\ast
}(\varphi )|}$ states is finite). By finite model property, if a formula $%
\varphi $ is consistent, then $\varphi $ is satisfiable with respect to some
finite models. Conversely, if $\varphi $ is satisfiable with respect to some
finite models, then $\varphi $ is consistent.

\textbf{Proposition 4} (Decidability property of $BPICTL$). The provability
problem for $BPICTL$ is decidable.

$Proof.$ Since $\varphi $ is provable in $BPICTL$ iff $\lnot \varphi $ is
not $BPICTL$-consistent, we can simply check if $\lnot \varphi $ is $BPICTL$%
-consistent. By the above discussion, there is a checking procedure. Hence
the provability problem for $BPICTL$ is decidable.

\section{Model Checking Algorithm for $BPICTL$}

In this section we give a model checking algorithm for $BPICTL.$ The model
checking problem for $BPICTL$ asks, given a model $M$ and a $BPICTL$ formula 
$\varphi ,$ for the set of states in $S$ that satisfy $\varphi .$ In the
following, we denote the desired set of states by $Eval(M,\varphi ),$ where $%
M=(S,\Pi ,\pi ,R_{a}^{B},R_{a}^{P},R_{a}^{I},R_{a}^{X},D_{a})$. The
algorithm will operate by labeling each state $s$ with the set $label(s)$ of
subformulas of $\varphi $ which are true in $s$. Initially, $label(s)$ is $%
\varnothing $. The algorithm goes through a series of stages. During the $i$%
th stage, subformulas with $i-1$ nested operators are processed. When a
subformula is processed, it is added to the labeling of each state in which
it is true. Once the algorithm terminates, we will have that $(M,s)\models
\varphi $ iff $\varphi \in label(s).$ Finally, the algorithm returns the set
of all $s$ labeled $\varphi .$ The case in which $\varphi =EG\theta $ is
based on the decomposition of the graph into strongly connected components.
A strongly connected component ($SCC$) $C$ is a maximal subgraph such that
every node in $C$ is reachable from every other node in $C$ along a directed
path entirely contained within $C$. $C$ is nontrivial iff either it has more
than one node or it contains one node with a self-loop.

Procedure $Eval(M,\varphi )$

\ \ case $\varphi $ is an atomic formula $p:$ if $\pi (s,p)=true$ then $%
label(s):=label(s)\cup \{p\};$

\ \ case $\varphi $ $=\lnot \theta :$ $T:=Eval(M,\theta );$ for all $s\notin
T$ do $label(s):=label(s)\cup \{\lnot \theta \};$

\ \ case $\varphi =\theta _{1}\wedge \theta _{2}:$ $T_{1}:=Eval(M,\theta
_{1});$ $T_{2}:=Eval(M,\theta _{2});$ for all $s\in T_{1}\cap T_{2}$ do $%
label(s):=label(s)\cup \{\theta _{1}\wedge \theta _{2}\};$

\ \ case $\varphi =\theta _{1}\vee \theta _{2}:$ $T_{1}:=Eval(M,\theta
_{1}); $ $T_{2}:=Eval(M,\theta _{2});$ for all $s\in T_{1}\cup T_{2}$ do $%
label(s):=label(s)\cup \{\theta _{1}\vee \theta _{2}\};$

\ \ case $\varphi =B_{a}\theta :T:=Eval(M,\theta );$ for all $s\in
Pre_{a}^{B}(M,T)$ do $label(s):=label(s)\cup \{B_{a}\theta \};$

\ \ case $\varphi =P_{a}\theta :T:=Eval(M,\theta );$ for all $s\in
Pre_{a}^{P}(M,T)$ do $label(s):=label(s)\cup \{P_{a}\theta \};$

\ \ case $\varphi =I_{a}\theta :T:=Eval(M,\theta );$ for all $s\in
Pre_{a}^{I}(M,T)$ do $label(s):=label(s)\cup \{I_{a}\theta \};$

\ \ case $\varphi =AX\theta :T:=Eval(M,\theta );$ for all $s\in
Pre_{a}^{AX}(M,T)$ do $label(s):=label(s)\cup \{AX\theta \};$

\ \ case $\varphi =EX\theta :T:=Eval(M,\theta );$ for all $s\in
Pre_{a}^{EX}(M,T)$ do $label(s):=label(s)\cup \{EX\theta \};$

\ \ case $\varphi =EF\theta :$

\ \ $\ \ T:=Eval(M,\theta );$

\ \ \ $\ $for all $s\in T$ do $label(s):=label(s)\cup \{EF\theta \};$

\ \ \ \ while $T\neq \varnothing $ do

\ \ \ \ \ \ choose $s\in T;$

\ \ \ \ \ \ $T:=T\backslash \{s\};$

\ \ \ \ \ \ for all $t$ such that $(t,s)\in R_{a}^{X}$ do

\ \ \ \ \ \ \ \ if $EF\theta \notin label(t)$ then

\ \ \ \ \ \ \ \ $\ \ label(t):=label(t)\cup \{EF\theta \};$

\ \ \ \ \ \ \ \ $\ \ T:=T\cup \{t\};$

\ \ \ \ \ \ \ \ end if;

\ \ \ \ \ \ end for all;

\ \ \ \ end while;

\ \ end case;

\ \ case $\varphi =EG\theta :$

$\ \ \ \ S^{\prime }:=Eval(M,\theta );$

\ $\ \ \ SCC:=\{C$ \TEXTsymbol{\vert} $C$ is a nontrivial strongly connected
component of $S^{\prime }$ with respect to $R_{a}^{X}$\};

$\ \ \ \ T:=\cup _{C\in SCC}\{s$ \TEXTsymbol{\vert} $s\in C\};$

\ \ \ \ for all $s\in T$ do $label(s):=label(s)\cup \{EG\theta \};$

\ \ \ \ while $T\neq \varnothing $ do

\ \ \ \ \ \ choose $s\in T;$

\ \ \ \ \ \ $T:=T\backslash \{s\};$

\ \ \ \ \ \ for all $t$ such that $t\in S^{\prime }$ and $(t,s)\in R_{a}^{X}$
do

\ \ \ \ \ \ \ \ if $EG\theta \notin label(t)$ then

$\ \ \ \ \ \ \ \ \ \ label(t):=label(t)\cup \{EG\theta \};$

\ \ \ \ \ \ \ \ $\ \ T:=T\cup \{t\};$

\ \ \ \ \ \ \ \ end if;

\ \ \ \ \ \ end for all;

\ \ \ \ end while;

\ \ end case;

\ \ case $\varphi =E\theta _{1}U\theta _{2}:$

\ \ $\ \ T:=Eval(M,\theta _{2});$

\ \ \ $\ $for all $s\in T$ do $label(s):=label(s)\cup \{E\theta _{1}U\theta
_{2}\};$

\ \ \ \ while $T\neq \varnothing $ do

\ \ \ \ \ \ choose $s\in T;$

\ \ \ \ \ \ $T:=T\backslash \{s\};$

\ \ \ \ \ \ for all $t$ such that $(t,s)\in R_{a}^{X}$ do

\ \ \ \ \ \ \ \ if $E\theta _{1}U\theta _{2}\notin label(t)$ and $t\in
Eval(M,\theta _{1})$ then

\ \ \ \ \ \ \ \ $\ \ label(t):=label(t)\cup \{E\theta _{1}U\theta _{2}\};$

\ \ \ \ \ \ \ \ $\ \ T:=T\cup \{t\};$

\ \ \ \ \ \ \ \ end if;

\ \ \ \ \ \ end for all;

\ \ \ \ end while;

\ \ end case;

$\ \ Eval(M,\varphi ):=\{s$ $|$ $\varphi \in label(s)\};$

\ \ return $Eval(M,\varphi )$

The algorithm uses the following primitive operations:

(1) The function $Pre_{a}^{B},$ when given a model $M$ and a set $\rho
\subseteq S$ of states, returns the set of states $s$ such that from $s$ all
next states lie in $\rho $ through $R_{a}^{B}.$ Formally, $%
Pre_{a}^{B}(M,\rho )=\{s$ $|$ $s\in S$ such that $\forall t.(s,t)\in
R_{a}^{B}\Rightarrow t\in \rho \}.$

(2) The function $Pre_{a}^{P},$ when given a model $M$ and a set $\rho
\subseteq S$ of states, returns the set of states $s$ such that $R_{a}^{P}$
maps $s$ to a set included in $\rho .$ Formally, $Pre_{a}^{P}(M,\rho )=\{s$ $%
|$ $s\in S$ such that $\rho \in R_{a}^{P}(s)\}.$

(3) The function $Pre_{a}^{I},$ when given a model $M$ and a set $\rho
\subseteq S$ of states, returns the set of states $s$ such that $R_{a}^{I}$
maps $s$ to a set included in $\rho .$ Formally, $Pre_{a}^{I}(M,\rho )=\{s$ 
\TEXTsymbol{\vert} $s\in S$ such that $\rho \in R_{a}^{I}(s)\}.$

(4) The function $Pre_{a}^{AX},$ when given a model $M$ and a set $\rho
\subseteq S$ of states, returns the set of states $s$ such that from $s$ all
next states lie in $\rho $ through $R_{a}^{X}.$ Formally, $%
Pre_{a}^{AX}(M,\rho )=\{s$ $|$ $s\in S$ such that $\forall t.(s,t)\in
R_{a}^{X}\Rightarrow t\in \rho \}.$

(5) The function $Pre_{a}^{EX},$ when given a model $M$ and a set $\rho
\subseteq S$ of states, returns the set of states $s$ such that from $s$
there exists a next state in $\rho $ through $R_{a}^{X}.$ Formally, $%
Pre_{a}^{EX}(M,\rho )=\{s$ $|$ $s\in S$ such that $\exists t.(s,t)\in
R_{a}^{X}\wedge t\in \rho \}.$

(6) Union, intersection, difference, and inclusion test for state sets.

Partial correctness of the algorithm can be proved by induction on the
structure of the input formula $\varphi .$ Termination is guaranteed since
the state space $S$ is finite.

\textbf{Proposition} \textbf{5.} The algorithm given in the above terminates
and is correct, i.e., it returns the set of states in which the input
formula is satisfied.

\section{Conclusions}

There were several works in representing, reasoning and verifying mental and
temporal properties in multi-agent systems \cite%
{BB05,HW02,HW03,KLP04,KP06,MKMPPP25,OSP24,WFHP02}. But interaction relations
between mental modalities and temporal modalities are seldom studied in the
previous works. Moreover, the previous works usually adopt Kripke structure
for semantics of epistemic temporal logic, but Kripke structure is not
suitable for semantics of preference modality and intention modality. In
this paper, we present a mental temporal logic $BPICTL,$ which is a powerful
language for expressing complex properties of multi-agent system such as
belief modality, preference modality, intention modality and temporal
modality. Some new interaction relations between mental modalities and
temporal modalities are given in $BPICTL$. We present a neighbourhood
semantics and an inference system for $BPICTL.$ The soundness, completeness,
finite model property and decidability of $BPICTL$ are proved. To verify
multi-agent systems, we present a model checking algorithm for $BPICTL.$

\end{document}